\documentclass[twocolumn]{aastex63}

\accepted{\today}
\submitjournal{ApJ}

\usepackage{comment}
\usepackage{makecell}
\usepackage{amsmath}
\usepackage{lineno}
\usepackage{longtable}

\begin{document}

\author[0000-0002-8770-6764]{R\'eka K\"onyves-T\'oth}
\affiliation{Konkoly Observatory, Research Center for Astronomy and Earth Sciences, H-1121 Budapest Konkoly
Th. M. út 15-17., Hungary; MTA Centre of Excellence}
\affiliation{Department of Experimental Physics, Institute of Physics, University of Szeged, D\'om t\'er 9, Szeged, 6720 Hungary}
\affiliation{ELTE Eötvös Loránd University, Gothard Astrophysical Observatory, Szombathely, Hungary}

\author{B\'alint Seli}
\affiliation{Konkoly Observatory, Research Center for Astronomy and Earth Sciences, H-1121 Budapest Konkoly
Th. M. út 15-17., Hungary; MTA Centre of Excellence}
\affiliation{E\"otv\"os University, Department of Astronomy, Pf. 32, 1518 Budapest, Hungary}

\shorttitle{Post-maximum SLSNe-I}
\shortauthors{K\"onyves-T\'oth, R}

\correspondingauthor{R\'eka K\"onyves-T\'oth}
\email{konyvestoth.reka@csfk.org}

\graphicspath{{./}{}}

\title{Type W and Type 15bn subgroups of hydrogen-poor superluminous supernovae: pre-maximum diversity, post-maximum homogeneity?}

\begin{abstract}

In this study, we analyze the post-maximum spectra of a sample of 27 Type I superluminous supernovae (SLSNe-I) in order to search for physical differences between the so-called Type W and Type 15bn sub-types. This paper is a continuation of \citet{ktr21} and \citet{ktr22}. In the former, it was revealed that not all SLSNe-I show the W-shaped absorption feature between 4000 and 5000 \AA\ in the pre-maximum spectra, and two new SLSN-subgroups were disclosed: Type W, where the W-shaped feature is present, and Type 15bn, where it is missing. In the latter, it was shown that the pre-maximum photosphere of Type W SLSNe-I tend to be hotter compared to Type 15bn objects, and they are different regarding their ion composition, their early light curves and their geometry as well. For completeness, post-maximum data are analyzed in this paper. It is concluded that in terms of photospheric temperature and velocity, Type W and Type 15bn SLSNe decrease to a similar value by the post-maximum phases, and their pseudo-nebular spectra are nearly uniform.  Pseudo-equivalent width calculations show that the pEW of the wavelength range between  4166 and 5266 \AA\ evolve differently in case of the two sub-types, while the other parts of the spectra seem to evolve similarly. It was found that the host galaxies of the studied objects do not differ significantly in their star formation rate, morphology, stellar mass and absolute brightness. The main difference behind the bimodality of Type W and Type 15bn SLSNe-I therefore is in their pre-maximum evolution.

\end{abstract}


\section{Introduction}\label{sec:intro}

The most luminous stellar explosions, called superluminous supernovae (SLSNe) were discovered in the beginning of the 21th century as intrinsically rare, but exceedingly bright supernovae occurring mostly in faint dwarf or irregular galaxies showing low metallicity, and high star formation rate  \citep[e.g.][]{chen13, Lunnan13, Lunnan14, Leloudas15, Angus16, japelj16, perley16, chen17,schulze18,hatsu18,nicholl21}. At first, they were defined as supernovae having an absolute brightness of $M < -21$ at the peak of the light curve (LC) in all bands of the optical wavelengths \citep[see e.g.][]{quimby11,galyam12,nicholl15,galyam19,nicholl21}. However, later it was found that there are some lower luminosity events (showing $\sim$-19.8 mag in the maximum) that can be spectroscopically classified as SLSNe as well \citep[see e.g.][]{decia18,quimby18,galyam19}. Therefore rather than using the magnitude cut, nowadays SLSNe are classified by their unique spectra.

Spectroscopically, they can be divided empirically into the hydrogen-poor Type I and the hydrogen-rich Type II sub-types. The physical difference lies in the progenitor being stripped by stellar winds/binary interaction before explosion, or having an envelope of H preserved until the explosion. 

This paper focuses on the H-poor group of SLSNe (SLSNe-I), where two additional subgroups were created: Fast evolving SLSNe-I having a light curve rise-time of $\sim$30 days and steep photospheric velocity decline after the maximum, and Slow evolving objects showing a slower rise ($\sim$50 days), and nearly constant photospheric velocities \citep[e.g.][]{inserra17,inserra18,quimby18,inserra19,ktr21}. However, \citet{ktr21} found that the classification into Fast and Slow groups may be different photometrically and spectroscopically: while the spectroscopically fast evolving SLSNe-I in their sample showed fast rise in the LC, they revealed that spectroscopically slow SLSNe may show either fast or slow LC rise-times. Therefore, along the Fast and Slow attribute, "photometrically" and "spectroscopically" terms will be used in the followings to avoid confusion. 

Apart from the extreme brightness, the spectra of SLSNe-I are unique as well: they tend to show a steep blue continuum, and strongly blended, broad absorption features in the photospheric phase. The most commonly identified ions besides the widely present O II are C II, Si II and Fe III \citep[see e.g.][]{quimby11,lelo12,nicholl13,inserra13,mazzali16,nicholl16,liu17,quimby18,kumar20}. In a lot of cases, the spectra of SLSNe-I are similar to the spectra of normal or broad-line Ic SNe, with a  $\sim$30 day delay compared to them \citep[see e.g.][]{pastor10,inserra13,nicholl13,blanchard19}. Therefore according to \citet{nicholl21}, SLSNe-I can be called as "ultra-hot" Type Ic supernovae with considerably higher absolute brightness and slower evolution. Recently it was found that SN~2020wnt, a peculiar object falls into the continuum between Type Ic SNe and SLSNe-I \citep{gutierrez22,tinyanont22}.

A notable difference between normal Ic SNe and SLSNe-I lies in the temperature of the photosphere: SLSNe-I tend to show hotter temperatures compared to normal Type Ic SNe. This can explain the blueness of the continuum and the presence of some highly ionized elements in the spectra of SLSNe-I in the early pre-maximum phases. The peculiar, W-shaped absorption blend appearing between 3900 and 4500 \AA\ in the pre-maximum spectra of SLSNe-I is widely used to distinguish between Ic SNe and SLSNe-I. This W-shaped feature can be modeled using the lines of O II \citep[e.g.][]{quimby11,mazzali16,liu17}, or alternatively with the joint presence of some other elements, e.g. O III and C III \citep{quimby07, dessart19, galyam19b, ktr20-2}. Up to 2021, it was believed that this W-shaped blend plays a significant role in the spectrum formation of all known SLSNe-I, however, \citet{ktr21} and \citet{ktr22} revealed that there are some SLSNe-I that have similar spectrum evolution to SN~2015bn, where the W-shaped feature is missing, and the temperature of the photosphere is lower than 12000 K. Motivated by this finding, two new subgroups of SLSNe-I were created, the Type W and the Type 15bn SLSNe-I. \citet{ktr21} found that Type W SLSNe-I represent themselves in Fast and Slow sub-types in terms of both photometry and spectroscopy, while Type 15bn SLSNe-I occur only in the spectroscopically slow evolving group with a wide range of LC rise times (from a few 10 days to 150 days). It was concluded by \citet{ktr22} that one of the main physical difference between Type W and Type 15bn SLSNe-I can be found in the photospheric temperatures: Type W SLSNe-I tend to show hotter photosphere compared to Type 15bn objects. The two groups may differ in the presence (Type W) or absence (Type 15bn) of pre-peak bumps in the LC (see \citet{ktr22} for further details, and \citet{lelo12,nicholl15,smith16,vrees17,anderson18,lin20,chen21,fiore21,pursi22}, as original references), and in polarimetric respects as well (see \citet{ktr22} and \citet{lelo15b, inserra16,cikota18,lee19,maund21,pursi23}). Although the polarimetric sample of SLSNe-I is still sparse, the following hypothesis was created: Type W SLSNe-I may originate from spherically symmetric progenitors and show null-polarization, while a two component model having a more symmetric outer layer and a more asymmetric inner layer \citep[see][]{inserra16} may be true for Type 15bn SLSNe-I that show increasing polarization.
To validate or confute this statement, the examination of a large set of polarimetric data will be needed in the future.  

Since \citet{ktr21} and \citet{ktr22} focused on the pre-maximum spectroscopic evolution of SLSNe-I, the main goal of the present paper is to search for possible empirical and physical differences and similarities between Type W and Type 15bn SLSNe-I in the post-maximum phases, and to find out if they become homogeneous after the disappearance of the commonly present W-shaped pre-maximum absorption feature. In Section \ref{sec:data} we show our sample containing 27 SLSNe-I together with the used post-maximum data-set, and describe our motivations to examine post-maximum spectra of these SLSNe-I besides the pre-maximum examination that was presented in \citet{ktr22}. In Section \ref{sec:disc} we show post-maximum spectrum modeling of 2 representative objects in the sample, discuss the evolution of the photospheric temperatures and velocities in the post-maximum phases, and derive pseudo-equivalent widths of 3 regions of the spectra. The properties of the host galaxies will be discussed as well. Finally, in \ref{sec:sum}, we summarize our results.

\section{Motivation and data}\label{sec:data}

This study continues the thread of two recent papers, \citet{ktr21} and \citet{ktr22}. In \citet{ktr21}, two new sub-types of SLSNe-I were disclosed regarding the presence/absence of the W-shaped O II feature in the pre-maximum spectra, while in \citet{ktr22}, physical reasons behind this bimodality were searched by modeling the available pre-maximum spectra of 28 SLSNe-I. Later, SN2017faf was removed from the sample due to the uncertain origin of this transient (see Section 2 \citet{ktr22}), thus 27 SLSNe-I remained in the sample.

In the followings, we describe the main conclusions of the two parent-papers, and list our motivations to examine the post-maximum spectra of the SLSNe-I in our sample. 

\subsection{Two new subgroups of SLSNe-I and empirical differences between them}

In \citet{ktr21}, we examined a sample of 28 SLSNe-I in order to find  a correlation between the evolution time-scales and the ejecta masses, and as a side product of this procedure, two new subgroups of SLSNe-I were disclosed. Data of the studied objects were downloaded from he Open Supernova Catalog (OSC)\footnote{https://sne.space/} \citep{guill17}. The details of the sample selection together with the basic data and references of the 28 SLSNe-I are listed in Table 1 and Section 2 of \citet{ktr21}.

The analysis started with ejecta mass calculations from the following equations:
\begin{equation}
M_{\rm ej} ~=~ 4 \pi \frac {c}{\kappa} v_{ \rm phot} t_{\rm rise}^2, 
\end{equation}
where $M_{\rm ej}$ is the ejecta mass of the SLSNe-I, $v_{\rm phot}$ is the photospheric velocity measured in the pre- or near-maximum phase and $t_{\rm rise}$ is the rise-time of the light-curve from the moment of the explosion to the date of the maximum light.
The rise-time of the studied SLSNe-I were derived from the dates of explosion and maximum light estimated by the publicly available MOSFit extrapolations in the Open Supernova Catalog. These equations show that in order to infer the ejecta masses, photospheric velocity estimates are crucial. 

\begin{figure*}
\centering
\includegraphics[width=12cm]{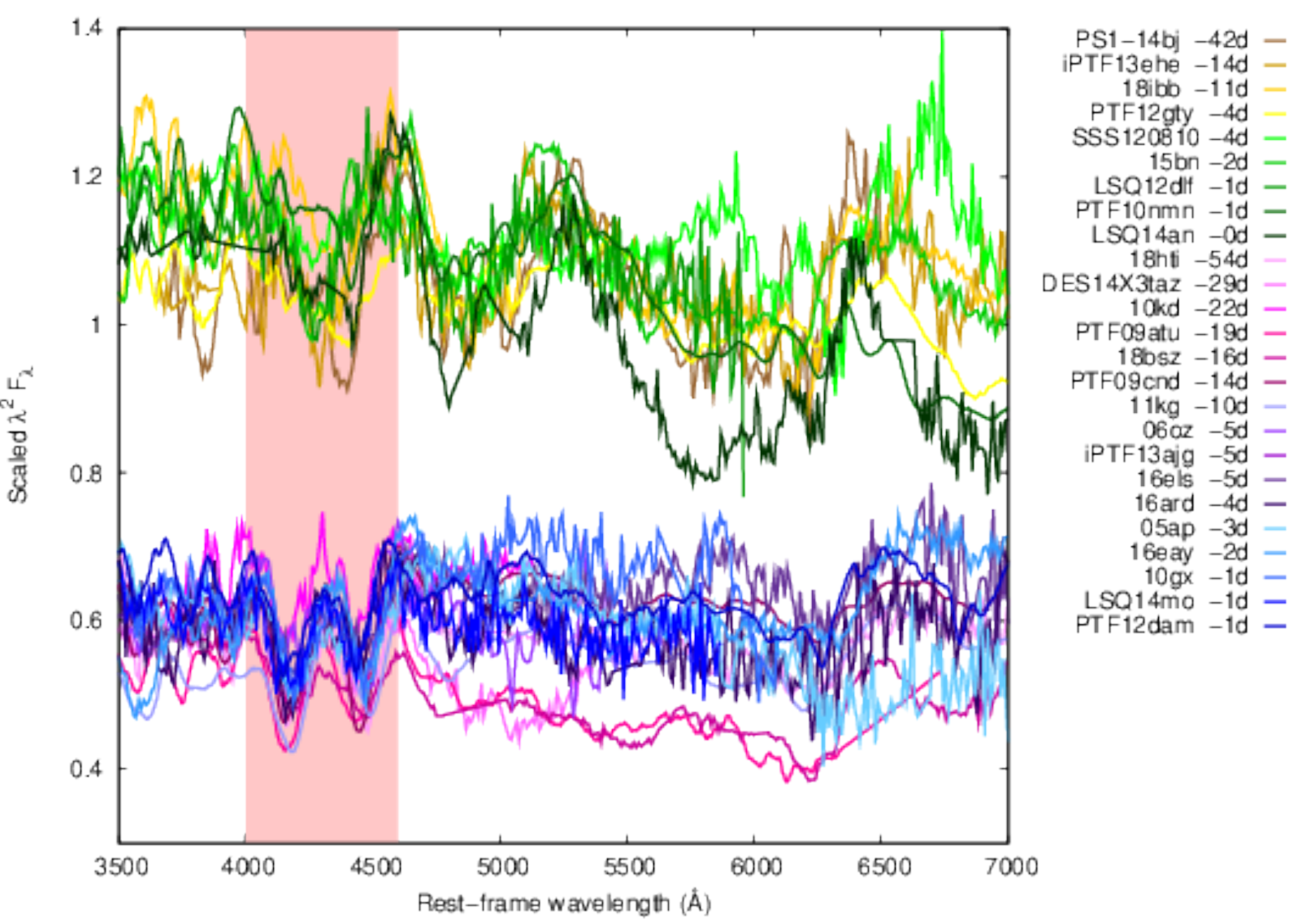}
\caption{Continuum normalized pre-maximum spectra of the studied Type W (on the bottom with purple-bluish colors) and Type 15bn SLSNe-I (on the top with orange-greenish colors) shifted vertically from each other, corrected for redshift and Milky Way extinction. The wavelength-range of the W-shaped O II feature is highlighted with a semi-transparent red rectangle to emphasize the difference between these two sub-types. A model obtained in SYN++ is plotted as well for Type W (red) and Type 15bn (blue) SLSNe-I to make the bimodality more visible. }
\label{fig:tipus}
\end{figure*}

However, modeling all available pre-maximum spectra would have been a time consuming method that leads far from the main goal, and therefore we combined the cross correlation technique with the SYN++ \citep{thomas11} spectrum modeling (see \citet{modjaz05} as well) in order to find a best-fit $v_{\rm phot}$ value for each object in a faster, but equally accurate way. The idea behind the usage of such method was the supposed presence of a W-shaped O II feature between $\sim$4000 and 5000 \AA\ in the pre/near-maximum spectra of the SLSNe-I in our sample that was believed to be omnipresent in SLSNe-I by some previous studies \citep[e.g.][]{quimby11,mazzali16,liu17}. Thus a series of O II models was built using SYN++ in the wavelength-range of the possibly present W-shaped absorption blend, and then cross correlated with the observed spectra to find out the best-fit photospheric velocity. However, in case of 9 SLSNe-I in the sample, the cross correlation resulted in physically senseless photospheric velocity values. Plotting together the spectra of these objects revealed the cause behind the discrepancy: they did not show the expected W-shaped feature between 4000 and 5000 \AA. Fortunately, the pre-maximum spectra of these 9 "outliers" were found to be similar to each other. This way, two new subgroups of SLSNe-I were defined: the objects showing the O II blend were called to be Type W SLSNe-I, while the "outliers" were named after a representative object, SN~2015bn as Type 15bn SLSNe-I. The difference between these  groups can be seen in Figure \ref{fig:tipus}, where the continuum normalized and redshift- and extinction corrected pre-maximum spectra of the studied SLSNe-I are plotted. The members of the two groups are shifted vertically from each other for clarification (see also Figure 1 in \citet{ktr22} that shows the difference between the two new sub-groups as well).

Apart from the distinct pre-maximum spectra, some additional differences between Type W and Type 15bn SLSNe-I were revealed in \citet{ktr21}. Studying the photometric and spectroscopic evolution time-scales, it was found that Type 15bn SLSNe-I show a roughly constant $v_{\rm phot}$ evolution between the pre-maximum phases and +30 rest-frame days after maximum. On the other hand, Type W objects were present in both the high, and the low velocity gradient group. It was shown as well that photometric and spectroscopic evolution is not necessarily the same: there are some Type 15bn objects that show a fast photometric evolution together with a slow spectroscopic evolution, while Type W SLSNe-I seem to be fast evolving both photometrically and spectroscopically (see Section 6 of \citet{ktr21} for further details).

\subsection{Possible physical differences between Type W and Type 15bn SLSNe-I}

In \citet{ktr22}, our main goal was to search for physical differences between Type W and Type 15bn SLSNe-I among the empirical ones. For this reason, detailed spectrum modeling of the available pre-maximum spectra was carried out using SYN++. In this code, one has to fit some global parameters, referring to the overall model spectrum, such as the photospheric temperature ($T_{\rm phot}$) and photospheric velocity ($v_{\rm phot}$), while other parameters code the lines of the individual elements (e.g. the optical depth, the width of the line-forming region, the feature width velocity, and the excitation temperature). The possibilities and constraints of the usage of SYN++, together with the uncertainties of the fitted $T_{\rm phot}$ values are discussed in Section 3.1 of \citet{ktr22}. 

Modeling the pre-maximum spectra of the studied objects revealed that Type 15bn SLSNe-I tend to show $T_{\rm phot} <$  12000 K, while Type W SLSNe-I are globally hotter in the pre-maximum phases having $T_{\rm phot} >$  12000 K. Accordingly, the chemical composition of the members of the two groups was found to be slightly different from each other. The C II and C III lines played a dominant role in the spectra of both Type W and Type 15bn SLSNe-I that implies the existence of a nearly pure carbon-oxygen envelope in these objects without the presence of large amount of heavier elements mixed up from the inner layers of the ejecta. Consistently with previous studies of SLSNe-I, oxygen plays a large role in the pre-maximum spectrum-formation of all studied objects, however, Type W SLSNe-I can be modeled using the joint presence of O I and O II, while Type 15bn objects show only O I. The most probable cause of this is the excitation of the O II lines that requires $T_{\rm phot} >$  12000 K \citep[e.g.][]{inserra19}. The borderline of $T_{\rm phot} =$  12000 K that was set to separate Type W and Type 15bn SLSNe-I is consistent with the other identified ions in the spectra. The Si and Fe lines were found to be significant and common in the spectra of both groups agreeing \citet{hatano99}, who showed that these ions form in a wide range of photospheric temperatures. On the contrary, Mg II and Ca II features require lower photospheric temperatures, and therefore they tend to be  missing from the spectra of Type W objects, while they are present in Type 15bn SLSNe-I. 

Although the size of our sample is still limited, and plenty of new observational data are required to prove or disprove unambiguously the following hypotheses, other possible physical differences may be found between the members of these two new subgroups. The first possibility lies in the "bumpiness" of the early light-curves, as was mentioned in Section \ref{sec:intro} as well. While the pre-maximum light-curve of Type 15bn SLSNe-I seem to evolve smoothly, without undulations, 6 of the Type W  SLSNe-I showed bumps in the early light-curve. Secondly, the geometry and the spectroscopic evolution is maybe connected as well. By the examination of the 5 SLSNe-I that have publicly available polarimetric data, we suspect that Type W SLSNe-I, SN~2018hti, LSQ14mo, and PTF12dam may originate from spherically symmetric explosions, given that the available polarimetric data showed null-polarization for these events. On the other hand, the polarization of SN~2015bn increases with time that can be explained with a two-layered model having  and an asymmetric inner layer and a spherically symmetric outer layer of the ejecta. It is important to mention though that SN~2018bsz, a Type W SLSN-I may show increasing polarization according to a single-epoch measurement. 

Maybe in the future, these statements can be examined in more details using a larger sample of SLSNe-I, and other photometric, spectroscopic and geometric differences may be found between Type W and Type 15bn sub-types. It is seen that the most notable difference between these groups lies in the presence/absence of the W-shaped O II feature before maximum, however, for completeness, the examination of the post-maximum data has some potential too: are these two groups similar to each other after the moment of the maximum, or they show other differences? Are there any other causes of the bimodality?

\subsection{Post-maximum data}

The main goal of this study is to examine the post-maximum data of the sample used in \citet{ktr21} and \citet{ktr22} in order to search for further reasons behind the distinction of the two new subgroups, Type W and Type 15bn SLSNe-I. Although the presence of the W-shaped absorption blend is the peculiarity of the pre-maximum phase in the spectroscopic evolution of SLSNe-I, it is important to clarify if the two groups remain different, or become similar in the post-maximum phases. This examination helps to reveal the physical causes of the bimodality. 

Here, we utilize 3 post-maximum spectra in case of each object in the sample: an "early post-maximum" spectrum taken between +1 and $\sim$+10 days rest-frame phase since maximum, a "later post-maximum" spectum taken between +10 and +35 days phase, and a "pseudo-nebular" spectrum using the definition of \citet{nicholl19}, so where some of the continuum is still present in the spectrum together with some strong nebular features (e.g. Ca H\&K $\lambda\lambda$3936,3968; Mg I] $\lambda$4571; [Fe II] $\lambda$5250; [O I] $\lambda\lambda\lambda$5577,6300,6363; [Ca II] $\lambda\lambda$7291,7323 and Ca II $\lambda\lambda\lambda$8498,8542,8662). Because of the lack of enough post-maximum data, the following objects were excluded from the sample: SN2006oz, DES14X3taz, LSQ14bdq and SN2018bsz from the Type W subgroup,  and SN2018ibb from the Type 15bn subgroup. 

\begin{figure*}
\centering
\includegraphics[width=8cm]{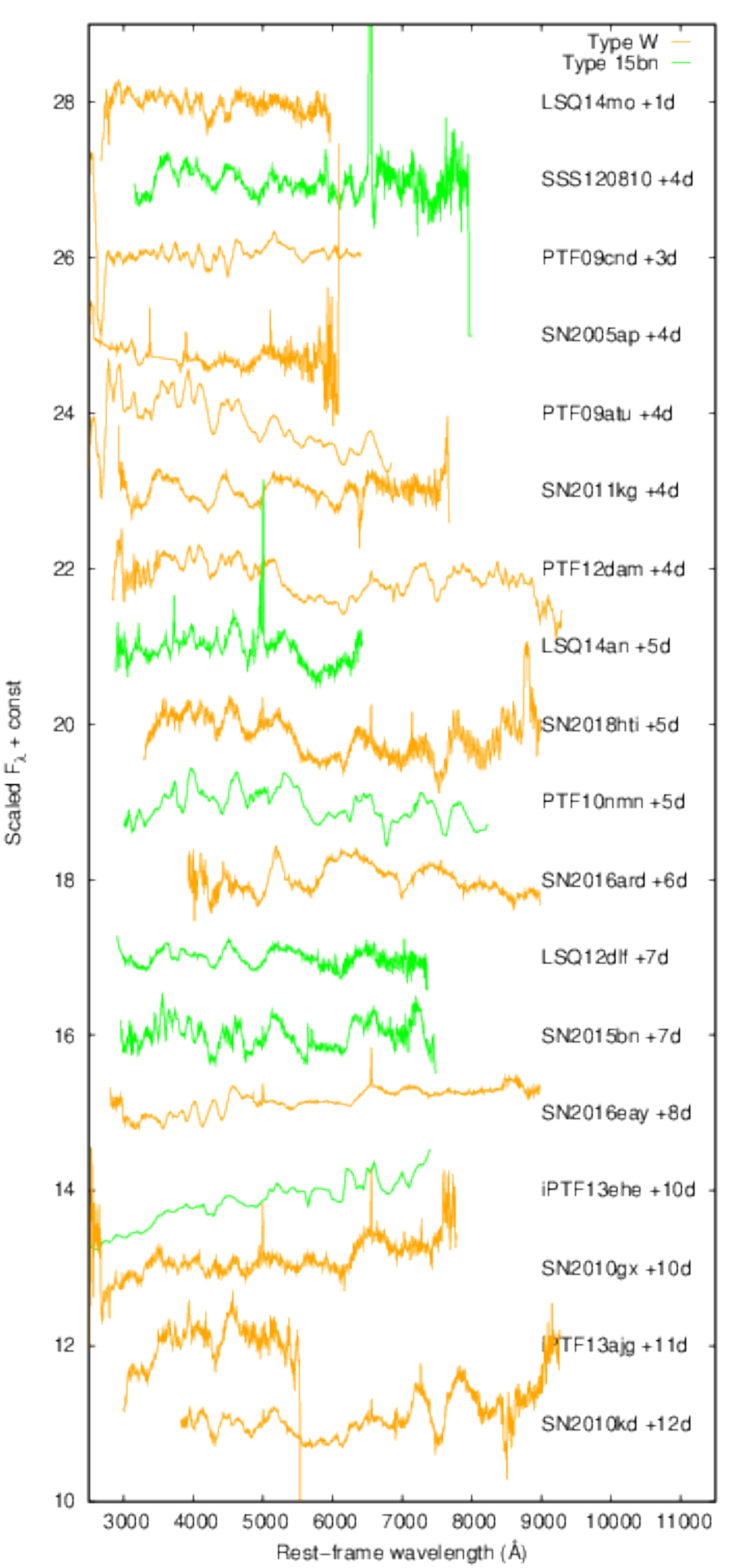}
\includegraphics[width=8cm]{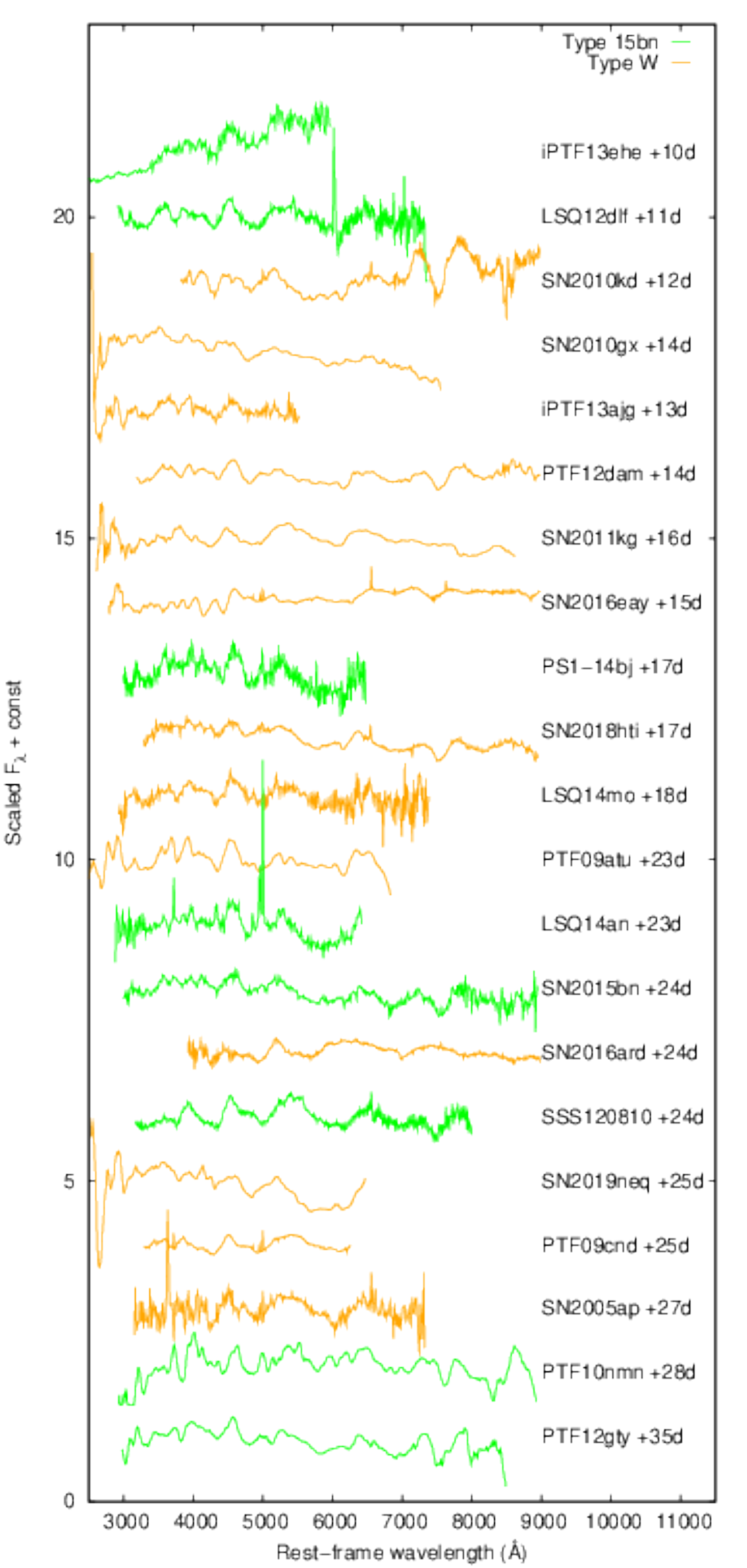}
\caption{Left panel: Continuum normalized early phase (between +1 and $\sim$+10 days) post-maximum spectra  of the studied SLSNe-I corrected for redshift and Milky Way extinction. Orange color denotes to Type W SLSNe-I, while Type 15bn objects are plotted with green. Right panel: Continuum normalized "later phase" (between +10 and +35 days) post-maximum spectra  of the studied SLSNe-I plotted with the same methodology and color coding as in the left panel. }
\label{fig:earlypost}
\end{figure*}

The left panel of Figure \ref{fig:earlypost} shows the redshift- and Milky Way extinction corrected early post-maximum spectra of the studied SLSNe-I arranged by phase. The spectra were continuum normalized before plotting using a fitted blackbody curve in order to make the features more visible. The temperatures of these black body curves can be found in Table \ref{tab:phases}. Type W SLSNe-I are plotted with orange, while green color denote the Type 15bn SLSNe-I. It is seen that in the early post-maximum phase, these spectra are quite diverse. There are some cases, where the W-shaped absorption blend is still clearly visible (e.g. LSQ14mo +1d, PTF09cnd +3d and SN2016eay +8d), while it is missing from the spectra of other Type W objects. The Type W SLSNe-I, where the W feature has already disappeared by this phase, are multitudinous as well: some of them are similar to the spectra of Type 15bn SLSNe (e.g. SN2011kg +4d, SN2018hti +5d), while others are completely different. On the other hand, the spectra of Type 15bn SLSNe-I seem to be mostly similar to each other in these early phases. The most probable cause of this phenomenon may be the photospheric temperature evolution of the objects. Type 15bn SLSNe-I show lower $T_{\rm phot}$ in the pre-maximum phases, which probably remains true in the post-maximum phases as well. On the contrary, Type W SLSNe-I have a larger range of $T_{\rm phot}$ in the pre-maximum phases, and therefore the change in the temperature is more drastic in some cases than in others. The evolution of the photospheric temperature and velocity will be discussed further in Section \ref{sec:tempvel}. 

In the right panel of Figure \ref{fig:earlypost}, the "later post-maximum" spectra of the studied SLSNe-I are plotted using the same color coding and methodology as in the left panel. By this phase, the spectra seem to become more or less similar to each other. One apparent exception is SN2016eay, where the W-shaped O II feature is still present. It is consistent with the very high photospheric temperature value, and slow $T_{\rm phot}$ evolution of this object (see Figure \ref{fig:grad} as well).
From all other Type W SLSNe-I, the W-shaped blend disappeared together with other pre-maximum features, and they became similar to Type 15bn SLSNe-I.

\begin{figure*}
\centering
\includegraphics[width=12cm]{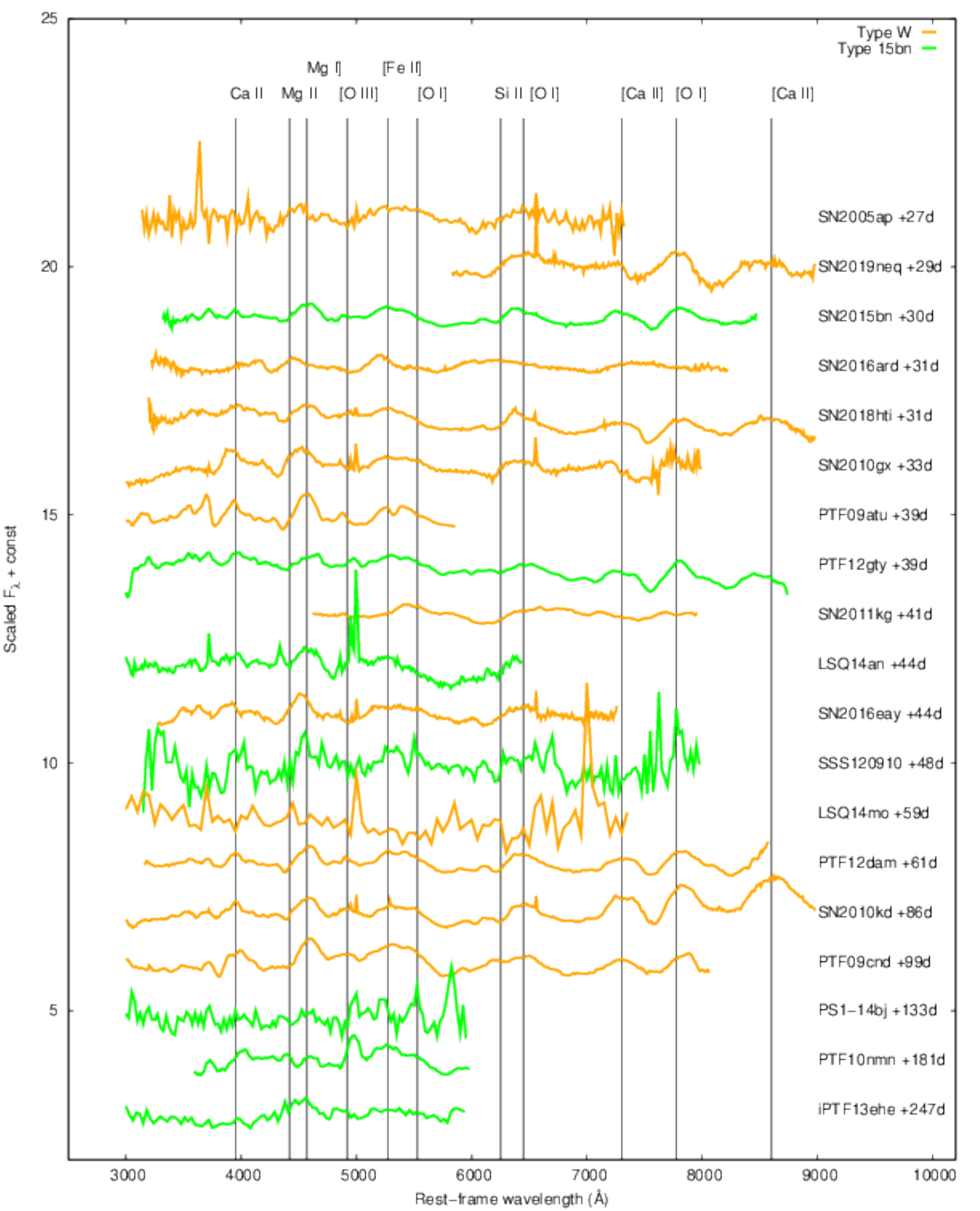}
\caption{Continuum normalized pseudo-nebular spectra  of the studied SLSNe-I plotted with the same methodology and color coding as in Figure \ref{fig:earlypost}. The places of the commonly present nebular lines are presented using black vertical lines.}
\label{fig:nebular}
\end{figure*}

\begin{figure*}
\centering
\includegraphics[width=12cm]{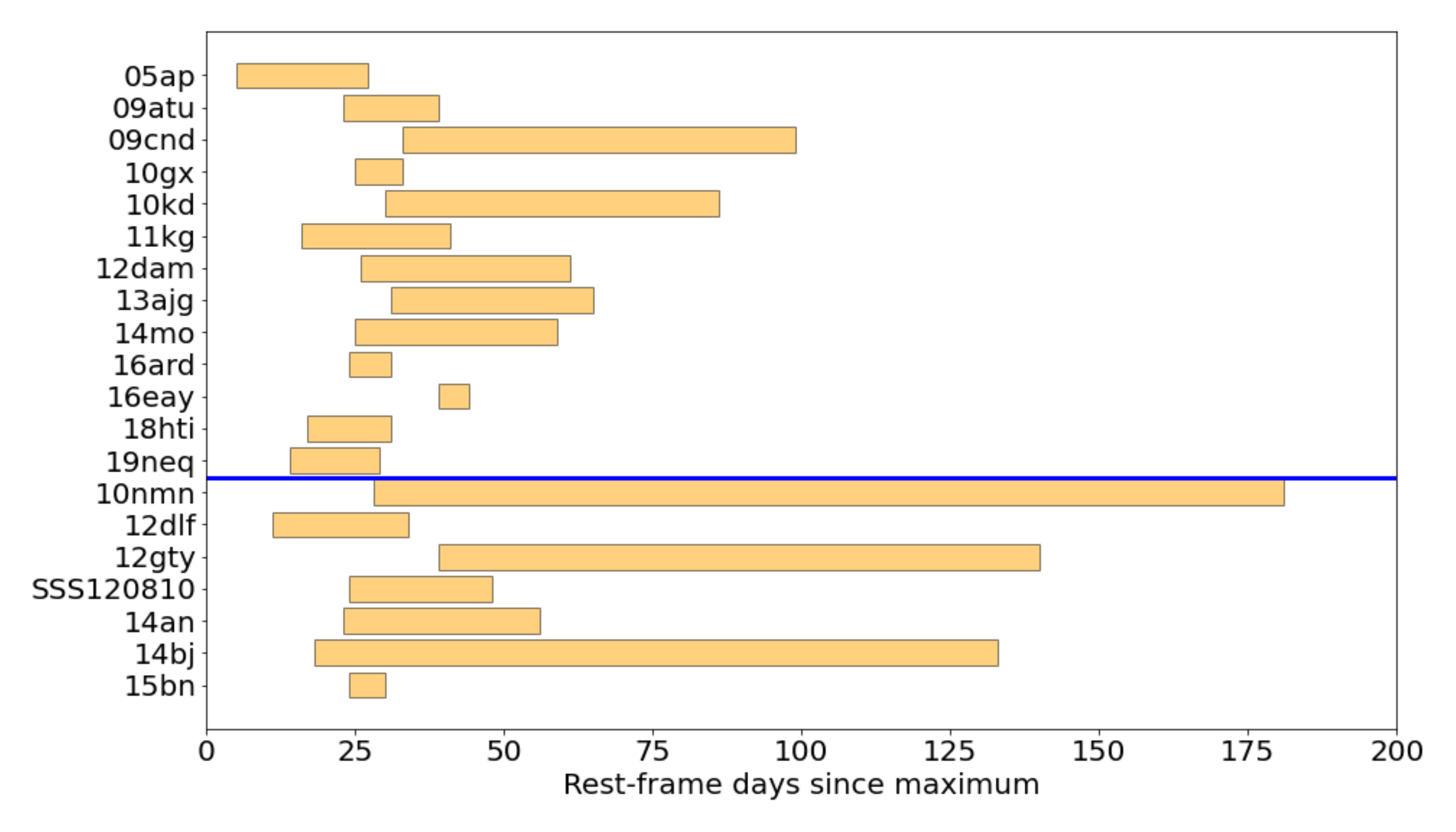}
\caption{The time interval of the transition from the photospheric to the nebular phase in case of each studied object  (orange rectangles). The left side of these rectangles shows the phase of the latest available post-maximum spectrum, while the right end of the rectangles denote to the epoch, when the spectra have already reached the pseudo-nebular phase.  Type W SLSNe-I are plotted above the blue horizontal line,  Type 15bn SLSNe-I can be seen below the blue line. }
\label{fig:atmenet}
\end{figure*}

Pseudo-nebular phase spectra of the SLSNe-I in our sample are plotted together in Figure \ref{fig:nebular} in the same way as in Figure \ref{fig:earlypost}. Consistently with \citet{nicholl19}, by the nebular phase, all SLSNe-I become similar to each other, and they cannot be separated into additional sub-classes. It seems to be true for the pseudo-nebular spectra as well, since no visible difference between Type W and Type 15bn SLSNe-I can be discovered any more. Figure \ref{fig:nebular} contains ion identifications as well, showing that most nebular features are present in the pseudo-nebular spectra of the studied objects. 

Here, another interesting question arises: when is the transition between the photospheric and the pseudo-nebular phase happening? Since we had only a few spectra in case of most studied SLSNe-I, only rough estimates could been made. Figure \ref{fig:atmenet} shows the period of this transition in case of each object with orange rectangles. The left side of these rectangles denote the phase of the latest available post-maximum spectrum, while the right end of the rectangles show the epoch, when the spectra have already reached the pseudo-nebular phase. Above the blue horizontal line, Type W SLSNe-I are plotted, while below the line, Type 15bn SLSNe-I are present. It is seen that the two groups cannot be specified with different transition times, so one cannot say that Type W SLSNe-I become pseudo-nebular faster than Type 15bn SLSNe-I. It contradicts the expectation, as that the photosheric velocity- and temperature evolution of Type W SLSNe-I seemed to be faster compared to Type 15bn SLSNe-I, so the earlier arrival to the pseudo-nebular phase would not have been unexpected. However, according to these data, we cannot distinguish between these groups by the time of the transition from the photospheric to the nebular phase. It can be read out from Figure \ref{fig:atmenet} as well that 9 SLSNe-I in the sample reached the pseudo-nebular phase before +30d phase, while most of them entered into this phase by $\sim$ +70d phase. 4 objects in the sample seem to reach the pseudo-nebular phase after +100 days: the Type W PTF09atu and the Type 15bn PTF10nmn, PTF12gty and PS1-14bj. However, because of the lack of high-cadence multi-epoch data of all objects in the sample, this statement must be treated cautiously, and will be examined in more details in the future.

\section{Discussion}\label{sec:disc}

\subsection{Spectrum modeling}

As was mentioned in Section \ref{sec:data}, Type W and Type 15bn SLSNe-I were found to be different during the pre-maximum phases by their photospheric temperatures, and accordingly their ion compositions (see Figure 8 in \citet{ktr22} for details). While the hotter photosphere of Type W SLSNe-I my allow the formation of the features of O II and other highly ionized elements, the pre-maximum spectra of Type 15bn SLSNe-I can be modeled using some lower ionization elements instead, e.g. Mg II and Ca II. Oxygen plays a significant role in case of the Type 15bn subgroup as well, however, only the neutral O I can be identified. Furthermore, in \citet{ktr21} it was revealed that all studied Type 15bn SLSNe-I show a slow spectroscopic evolution, while Type W SLSNe-I are present in both the fast and the slow evolving sub-types regarding  their photospheric velocity evolution. The presence of this pre-maximum bimodality raised the question: are the two groups different in their post-maximum phases too, as the temperature cools down in Type W SLSNe-I, and the W-shaped O II disappears? Or do they become similar to each other? Here, we seek the answer to these questions by modeling some representative post-maximum spectra of 2 SLSNe-I in the sample: Type W PTF12dam, and SN~2015bn itself. 

The cause for not modeling all available post-maximum spectra, beside of being very time consuming, can be seen in Figures \ref{fig:earlypost} and \ref{fig:nebular}. Although the "early post-maximum" spectra of the objects are quite diverse, even within the Type W and Type 15bn sub-groups, they become more similar by the "later post-maximum" phases, and almost uniform by the pseudo-nebular phases, consistently with \citet{nicholl19}. Therefore, two spectra were chosen for modeling in case of PTF12dam and SN~2015bn: an early post-maximum, and a later post-maximum spectrum. 

To model these spectra, the SYN++ code was utilized, similarly to the pre-maximum spectrum modeling shown in \citet{ktr22}. (See the possibilities and caveats of the spectrum modeling using SYN++ in Section 3.1 of \citet{ktr22}.)

\begin{figure*}
\centering
\includegraphics[width=8cm]{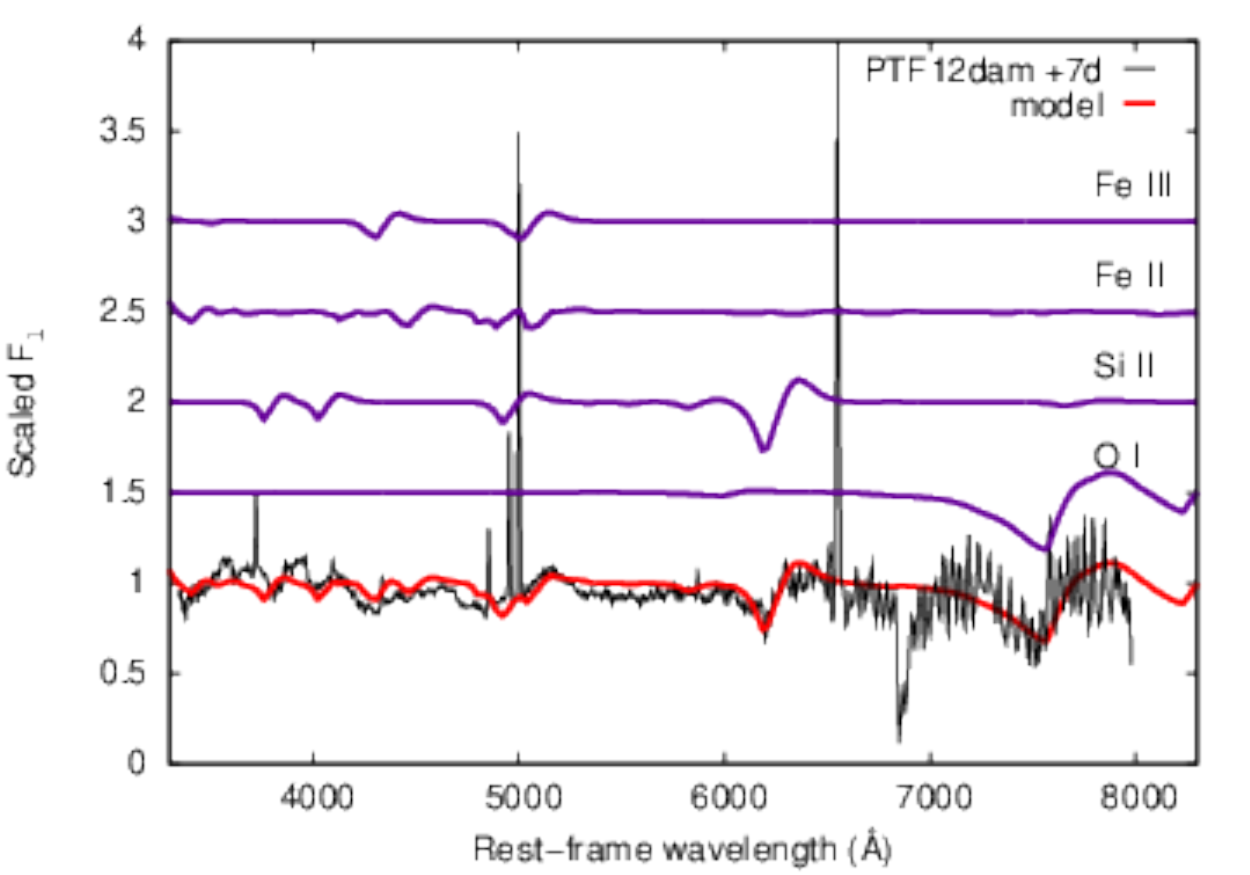}
\includegraphics[width=8cm]{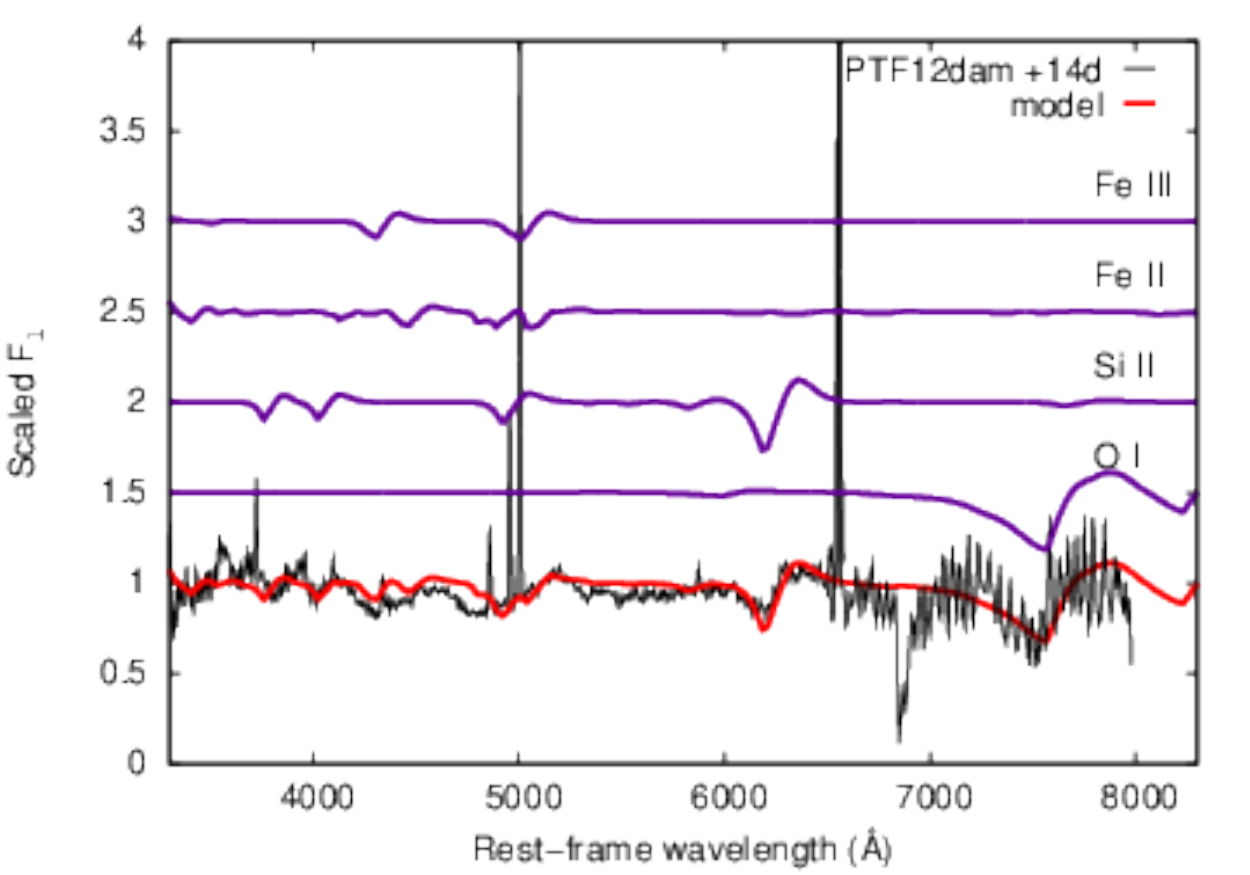}
\includegraphics[width=8cm]{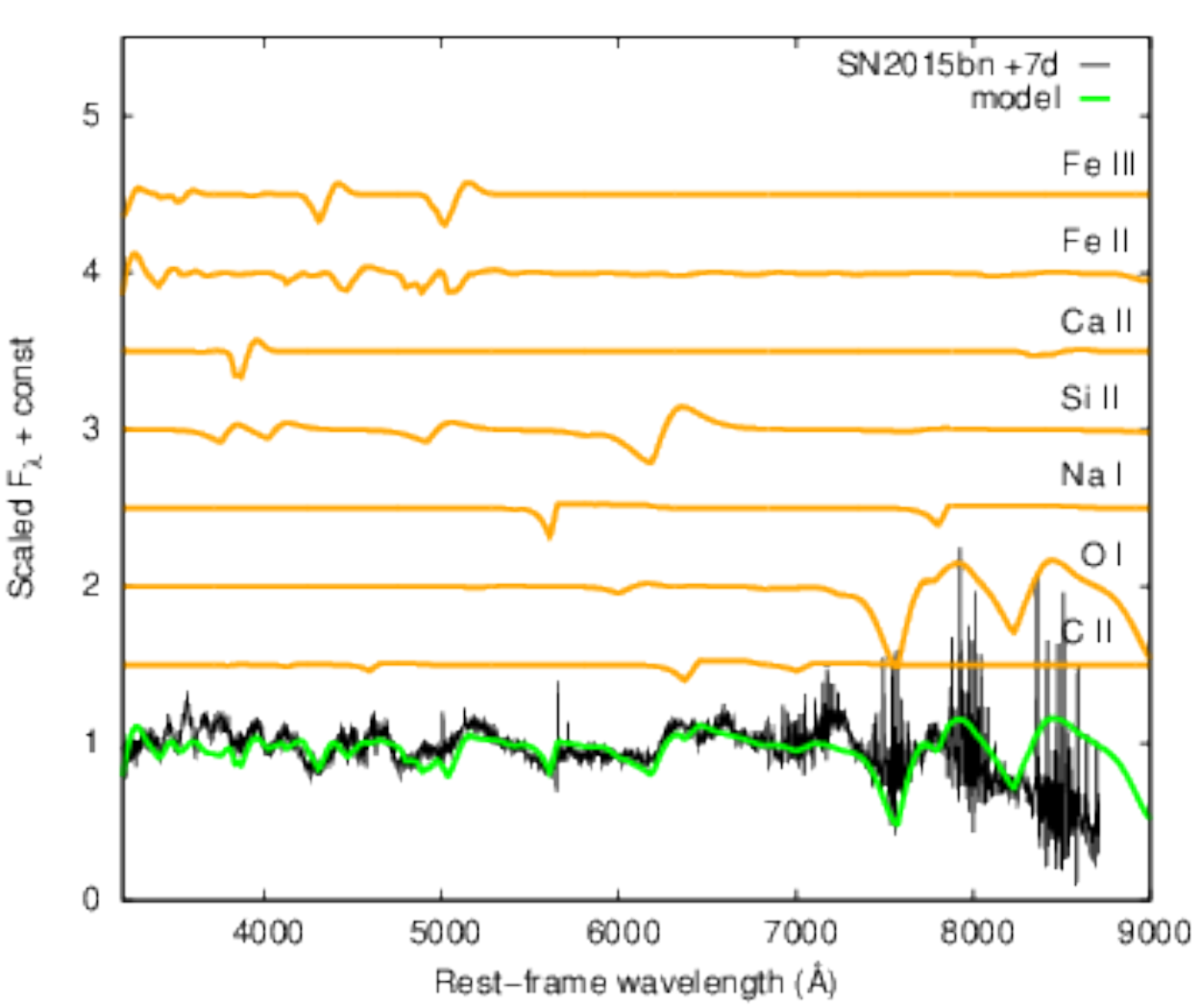}
\includegraphics[width=8cm]{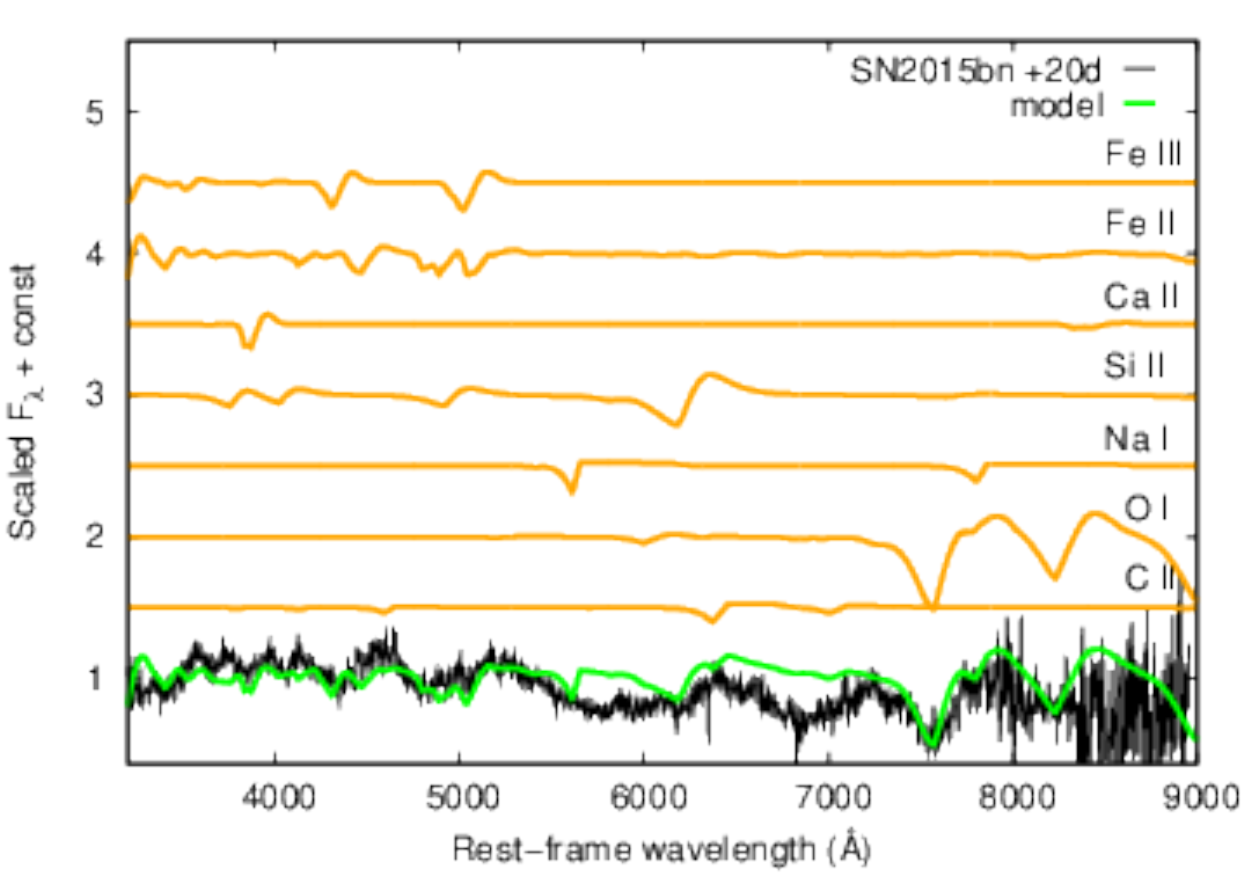}
\includegraphics[width=8cm]{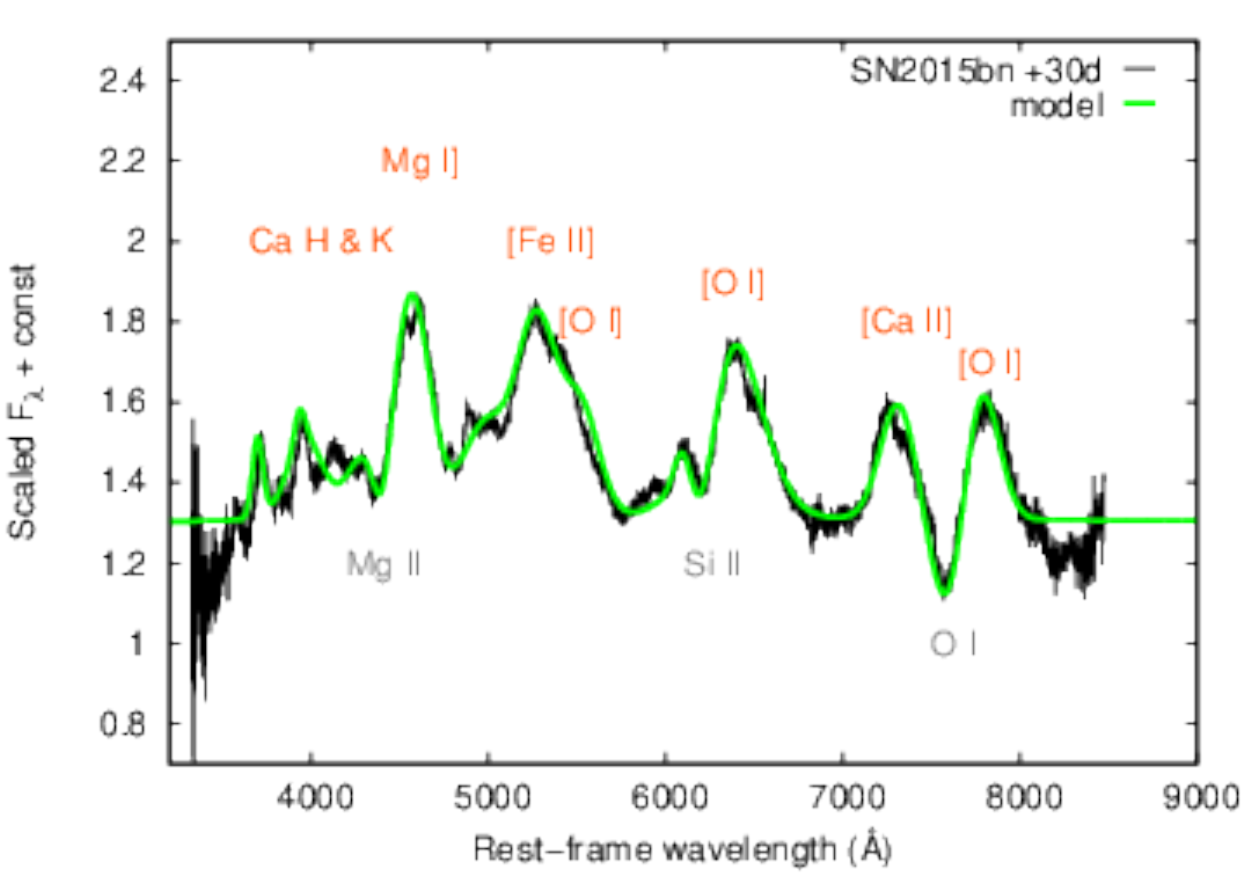}
\caption{Top left: Spectrum modeling of the "early post-maximum" (+7d phase) spectrum of PTF12dam, a Type W SLSN-I. The corrected and continuum normalized observed spectrum is plotted with black, while red color denote the best-fit model obtained in SYN++. Single ion contributions to the overall model spectrum are shown with purple lines shifted vertically from each other. Top right: Spectrum modeling of the "later post-maximum" (+14d phase) spectrum of PTF12dam with the same color coding as the top left panel. Middle left: spectrum modeling of the "early post-maximum" (+7d phase) spectrum of SN~2015bn, a Type 15bn SLSN-I (black line). The best-fit SYN++ model (green line), and the identified ions (orange lines) are plotted similarly to the top left panel. Middle right:  spectrum modeling of the "later post-maximum" (+20d phase) spectrum of SN~2015bn. Bottom: pseudo-nebular spectrum of SN~2015bn taken at +30d phase (black line) plotted together with the fitting with Gaussians (green line), and the identification of nebular emission features and absorption lines remained there from the photospheric phase. }
\label{fig:modeling}
\end{figure*}

The top panel of Figure \ref{fig:modeling} shows the best-fit model obtained for the "early post-maximum" phase (+7 days) and the "later post-maximum" phase (+14 days)  spectrum of PTF12dam with red line, while single ion contributions to the overall model are plotted with purple, shifted vertically from the observed spectrum. In the middle panel of Figure \ref{fig:modeling}, "early post-maximum" phase (+7 days) and the "later post-maximum" phase (+20 days)  spectrum of SN~2015bn are plotted (green lines) together with the observed spectra (black lines), and the single ion contributions (orange lines). From these figures, it becomes visible that by the post-maximum phase these objects are similar to each other in both observed epochs: O I, Si II, Fe II and Fe III lines are present in all cases strongly, while the best-fit models of SN~2015bn use C II, Na I and Ca II as well.  O II is not present any more in the spectrum of PTF12dam, because of the cooler photosphere compared to the pre-maximum epochs. The local parameters of the best-fit models for each epoch of the two modeled SLSNe-I can be found in Table \ref{tab:syn} 
in the Appendix, while the used $T_{\rm phot}$ and $v_{\rm phot}$ values are presented in Table \ref{tab:phases}.

Since the pseudo-nebular spectra of all studied objects are very similar to each other, only one of them, the +30 days phase spectrum of SN~2015bn is examined in more details. The bottom panel of Figure \ref{fig:modeling} shows the observed spectrum of SN~2015bn at +30d phase with black line, which was fitted with Gaussians (green line) of the typical nebular emission features (see the identifications in the plot): Ca H\&K $\lambda\lambda$3936,3968; Mg I] $\lambda$4571; [Fe II] $\lambda$5250; [O I] $\lambda\lambda\lambda$5577,6300,6363; [Ca II] $\lambda\lambda$7291,7323 and Ca II $\lambda\lambda\lambda$8498,8542,8662. Some absorption lines (Mg II, Si II and O I) are present in the spectrum as well. 

In connection with the identified elements, the change in the photospheric temperatures and velocities is apparent, and they will be discussed in details in Section \ref{sec:tempvel}.

\subsection{Photospheric velocity and temperature evolution}\label{sec:tempvel}

\begin{table*}
\footnotesize
\caption{The examined pre-maximum (Phase 1), "early post-maximum" (Phase 2), "later post-maximum" (Phase 3) and pseudo-nebular (Phase 4) epochs of each studied object together with the best-fit photospheric temperatures in the pre-maximum phase ($T_{\rm phot}$ 1), the "early post-maximum" phase ($T_{\rm phot}$ 2) and the "later post-maximum" phase ($T_{\rm phot}$ 3). Photospheric velocities calculated from the combination of SYN++ modeling and cross correlation technique can be found for the pre-maximum phase ($v_{\rm phot}$ 1) and the later post-maximum phase ($v_{\rm phot}$ 2). }
\label{tab:phases}
\centering
\begin{tabular}{lccccccccc}
\hline
\hline
SN & Phase 1 (a) & Phase 2 (b) & Phase 3 (b) & Phase 4 (b)  & $T_{\rm phot}$ 1 (a) & $T_{\rm phot}$ 2 (b) & $T_{\rm phot}$ 3 (b)  & $v_{\rm phot}$ 1 (a) & $v_{\rm phot}$ 2 (b) \\
  & (days) & (days) & (days) & (days) & (K) & (K) & (K) & (km s$^{-1}$) & (km s$^{-1}$) \\
\hline
\multicolumn{10}{c}{Type W} \\
\hline
SN~2005ap & -3 & 4 & 27 &27& 20000 & 13000 & 6500 & 23000 & 6000 \\
PTF09atu & -19 & 4 & 23 & 39&12000 & 12500 & 7000 & 10000 & 9000 \\
PTF09cnd & -14 & 3 & 25 & 99&14000 & 14000 & 9500 & 13000 & 9000 \\
SN~2010gx & -1 & 8 & 14 & 33&17000 & 15000 & 11500 & 20000 & 12000 \\
SN~2010kd & -22 & 12 & 12 &86& 15000 & 11500 & 11500 & 11000 & 9000 \\
SN~2011kg & -10 & 4 & 16 &41 &12000 & 9500 & 6000 & 11500 & 9000 \\
PTF12dam & -1 & 4 & 14 & 61&14000 & 12500 & 10000 & 12000 & 9000 \\
iPTF13ajg & -5 & 11 & 13 &--& 12000 & 12000 & 9500 & 10000 & 8000 \\
LSQ14mo & -1 & 1 & 18 & 59&12000 & 14000 & 7500 & 10000 & 6000 \\
SN~2016ard & -4 & 6 & 24 &31& 20000 & 10000 & 7000 & 14000 & 7000 \\
S2016eay &-2 & 8 &15&44& 20000 &17000& 17000& 20000& 8500 \\
SN~2018hti & -54 & 5 & 17 &--& 13000 & 14000 & 7000 & 15000 & 8000 \\
SN~2019neq & -4 & 24 & 24 &29& 14000 & 7000 & 7000 & 24000 & 8000 \\
\hline
\multicolumn{10}{c}{Type 15bn} \\
\hline
PTF10nmn & -1 & 5 & 28 &181& 8000 & 7000 & 7000 & 8000 & 6000 \\
PTF12gty & -4 & -- & 35 &39& 8500 & -- & 8500 & 8000 & 7500 \\
LSQ12dlf & -1 & 7 & 11 &--& 11000 & 9000 & 8000 & 15000 & 9500 \\
iPTF13ehe & -14 & 10 & 10 &274& 11000 & 8500 & 8500 & 10000 & 10000 \\
LSQ14an & -0 & 5 & 23 &44 &8000 & 9000 & 7000 & 8000 & 8000 \\
PS1-14bj & -42 & -- & 17 &133& 10000 & -- & 8000 & 10000 & 8000 \\
SN~2015bn & -17 & 7 & 20 &30 &12000 & 10500 & 9000 & 10000 & 8000 \\
SN~2018ibb (c) & -11 & 1 & -- & --&11000 & 10500 & -- & 8000 & -- \\
SSS120810 & -1 & 4 & 24 & 48&10000 & 9500 & 6000 & 10000 & 7000 \\
\hline
\end{tabular}
\tablecomments{a: adopted from \citet{ktr22}, b: examined/inferred in the present paper, c: SN~2018ibb was included in the photospheric temperature and velocity calculations because of its available pre-maximum and near maximum data.
}
\end{table*}

The next step is to calculate the photospheric velocities and temperatures of the studied objects, and to follow the evolution of these parameters from the pre-maximum to the post-maximum phases.

The post-maximum photospheric velocities were calculated by cross-correlating each "later post-maximum" spectra with the +20d spectrum of SN~2015bn. The "later post-maximum" phases were chosen instead of the "early post-maximum" phases, because by the later phases, the spectra of Type W and Type 15bn SLSNe-I become mostly homogeneous, so that the usage of the same template spectrum is reasonable. Besides, the $v_{\rm phot}$  between $\sim$0d and $\sim$+30d is a representation of the fastness/slowness of each object (see more details e.g. in
\citet{ktr21}). 

\begin{figure*}
\centering
\includegraphics[width=15cm]{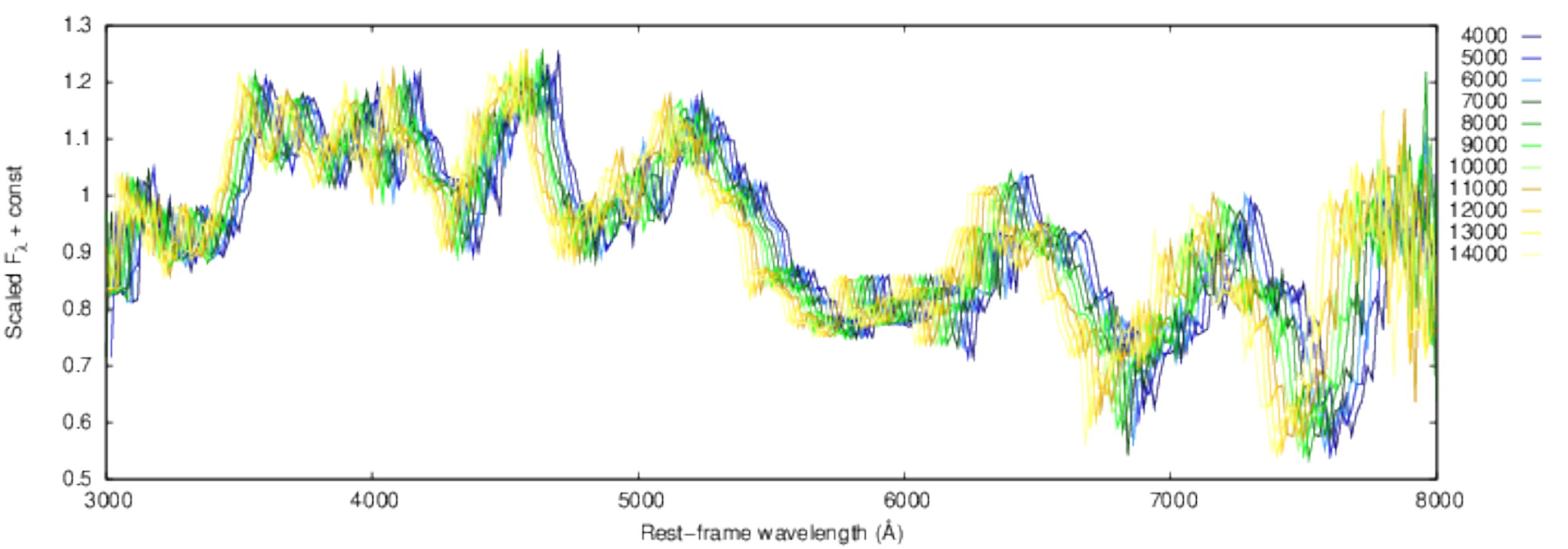}
\caption{The +20d phase redshift- and reddening corrected, continuum normalized spectrum of SN~2015bn ($v_{\rm phot}$ = 8000 km s$^{-1}$), and its different colored versions shifted in the velocity space between 4000 and 14000  km s$^{-1}$. They were used as templates in the cross correlation process to determine the photospheric velocities of all studied SLSNe-I. }
\label{fig:templet}
\end{figure*}

\begin{figure*}
\centering
\includegraphics[width=4cm]{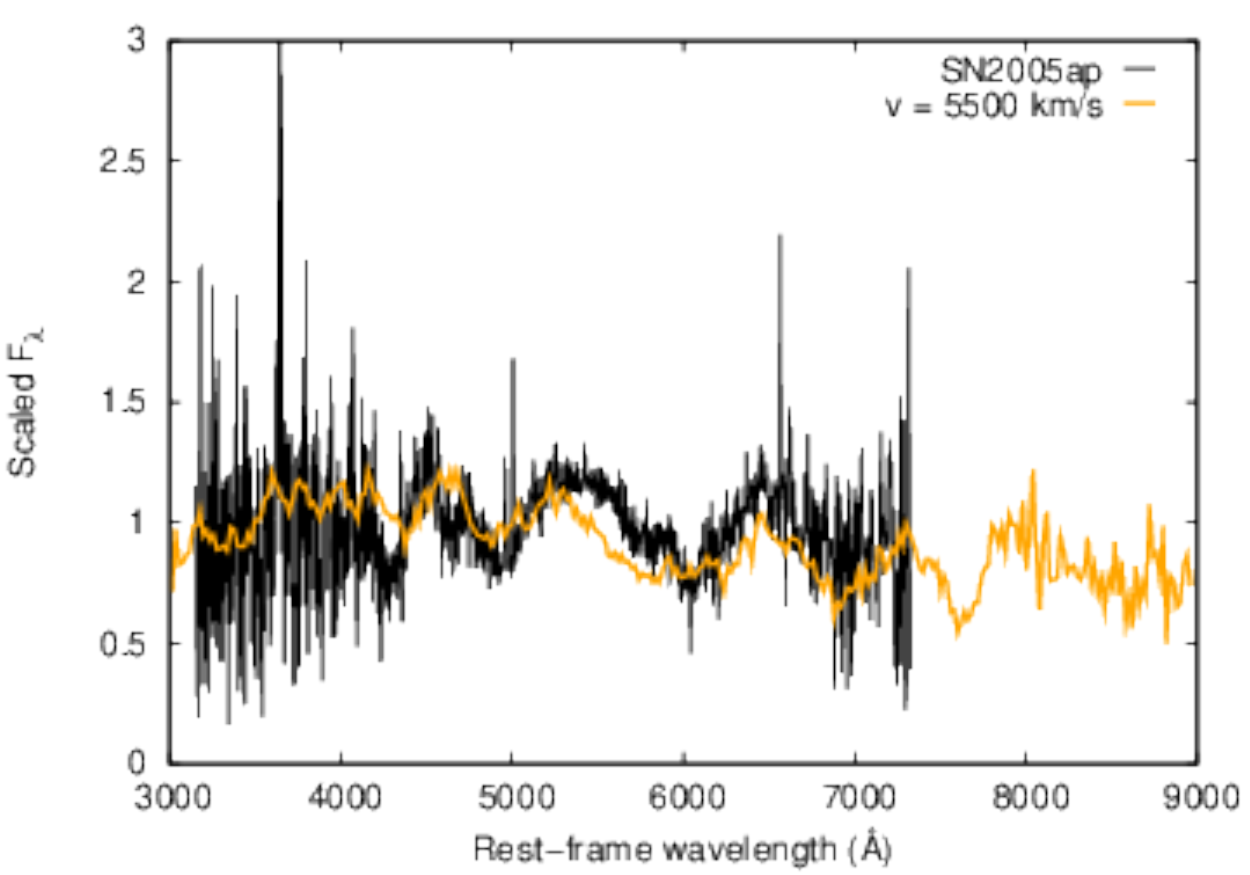}
\includegraphics[width=4cm]{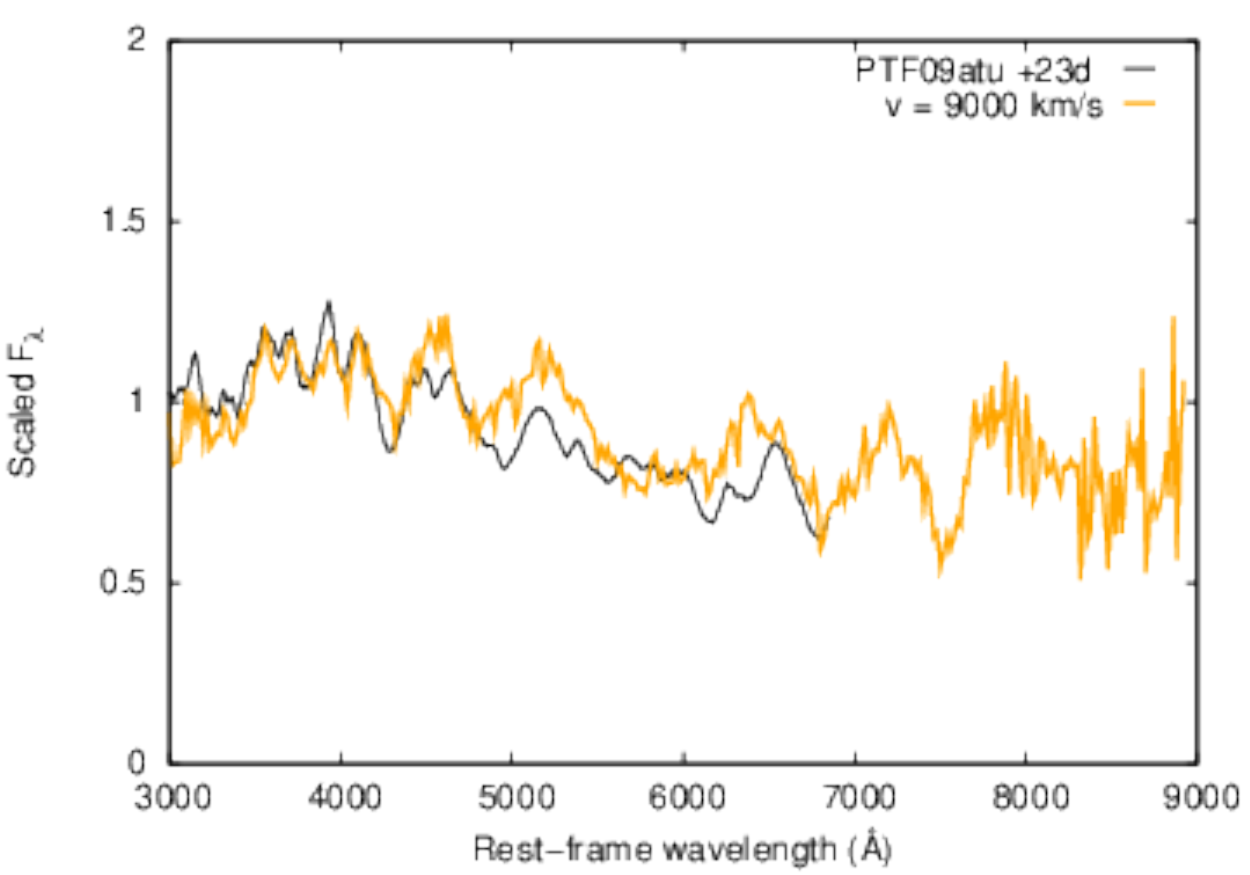}
\includegraphics[width=4cm]{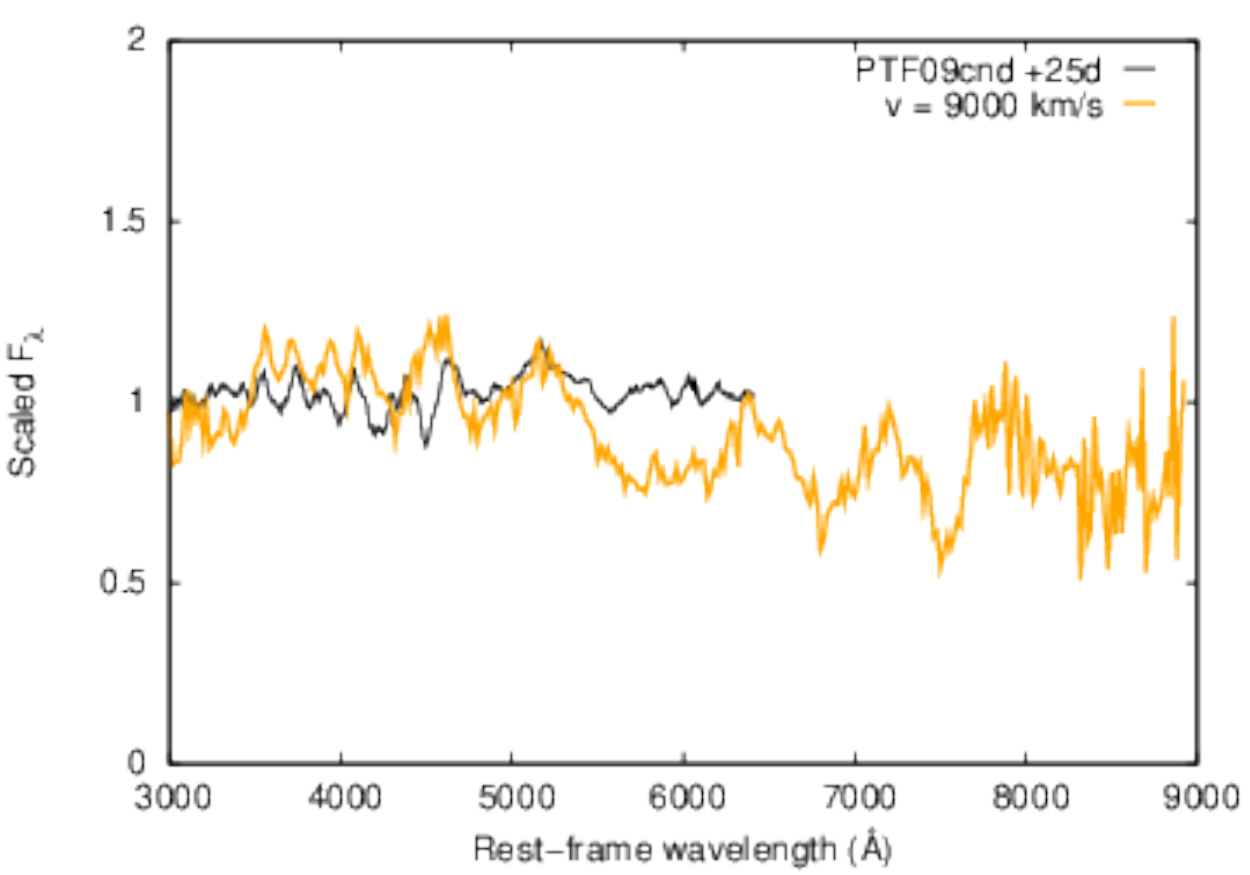}
\includegraphics[width=4cm]{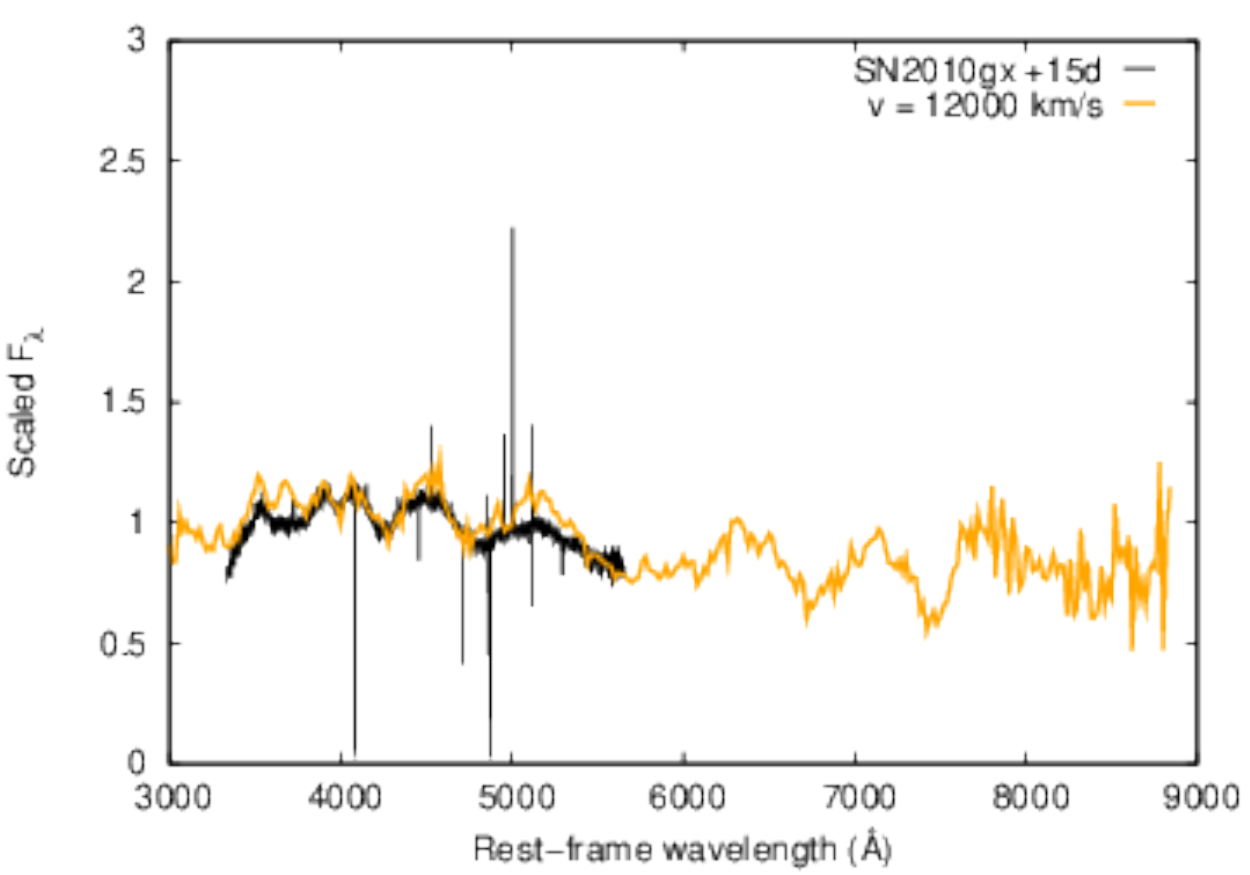}
\includegraphics[width=4cm]{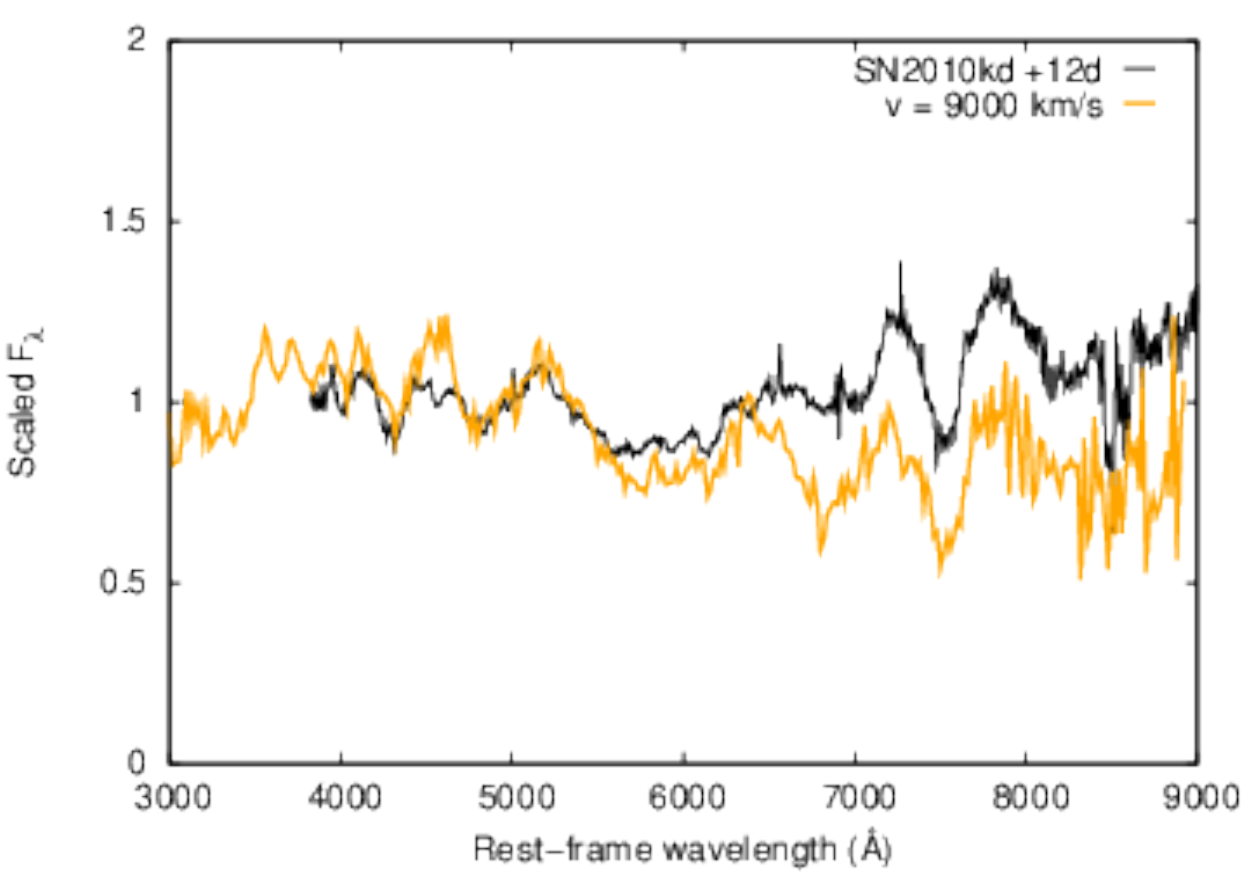}
\includegraphics[width=4cm]{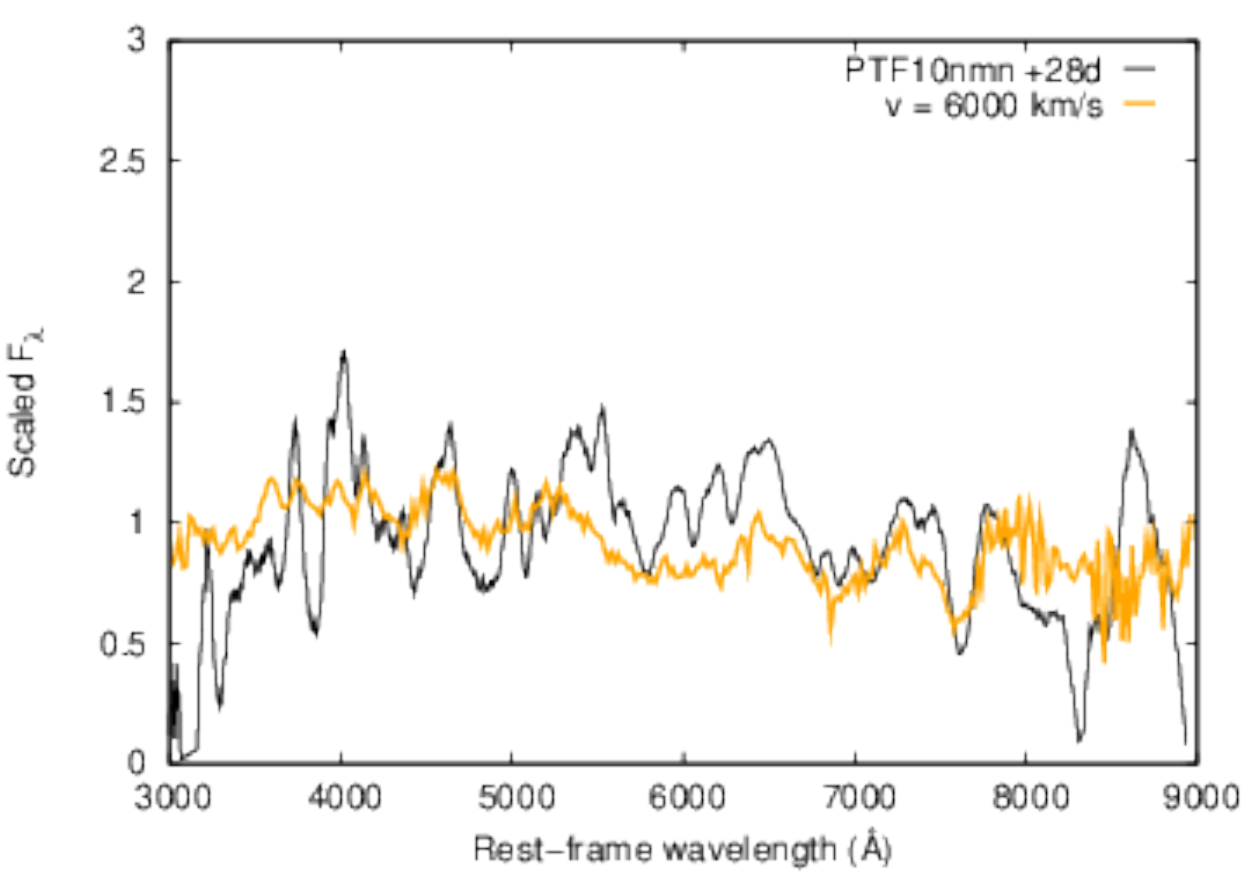}
\includegraphics[width=4cm]{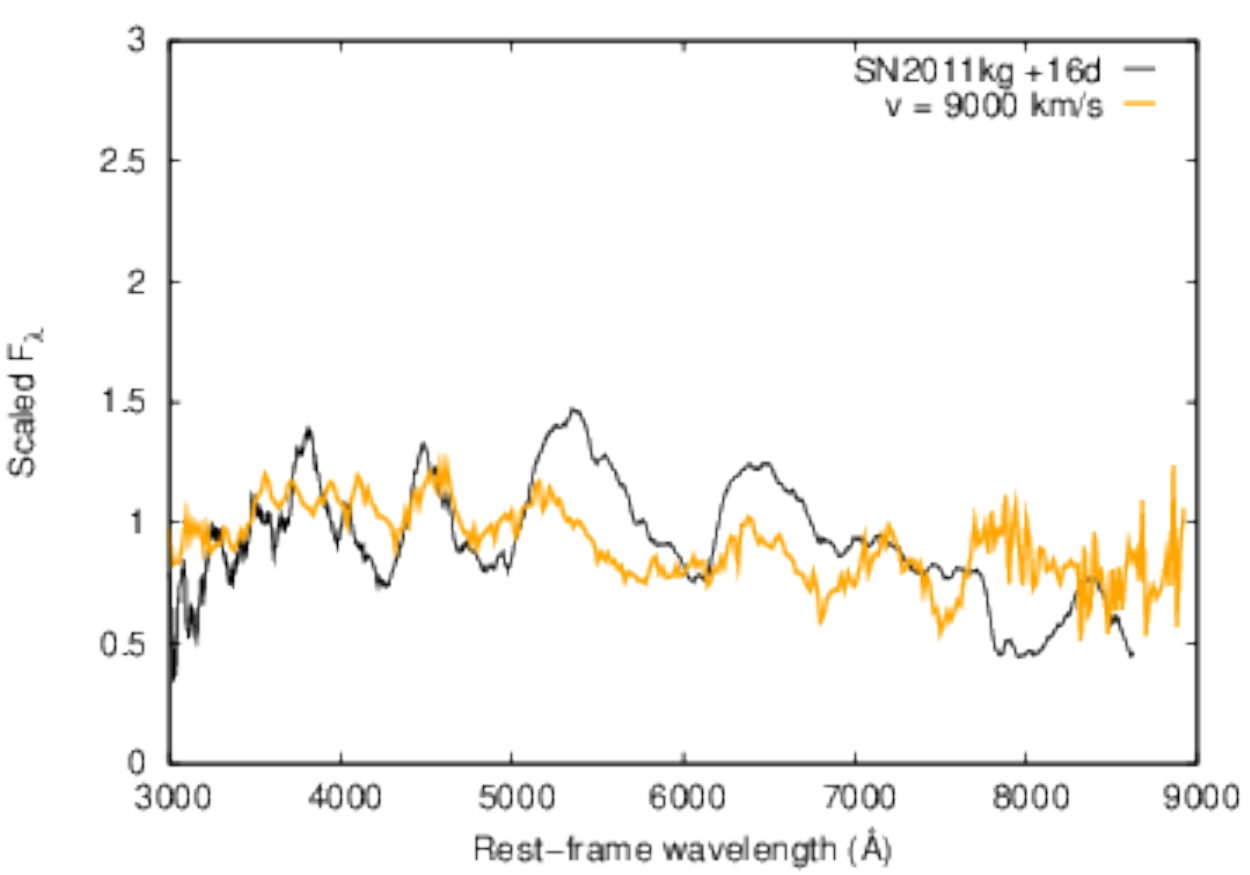}
\includegraphics[width=4cm]{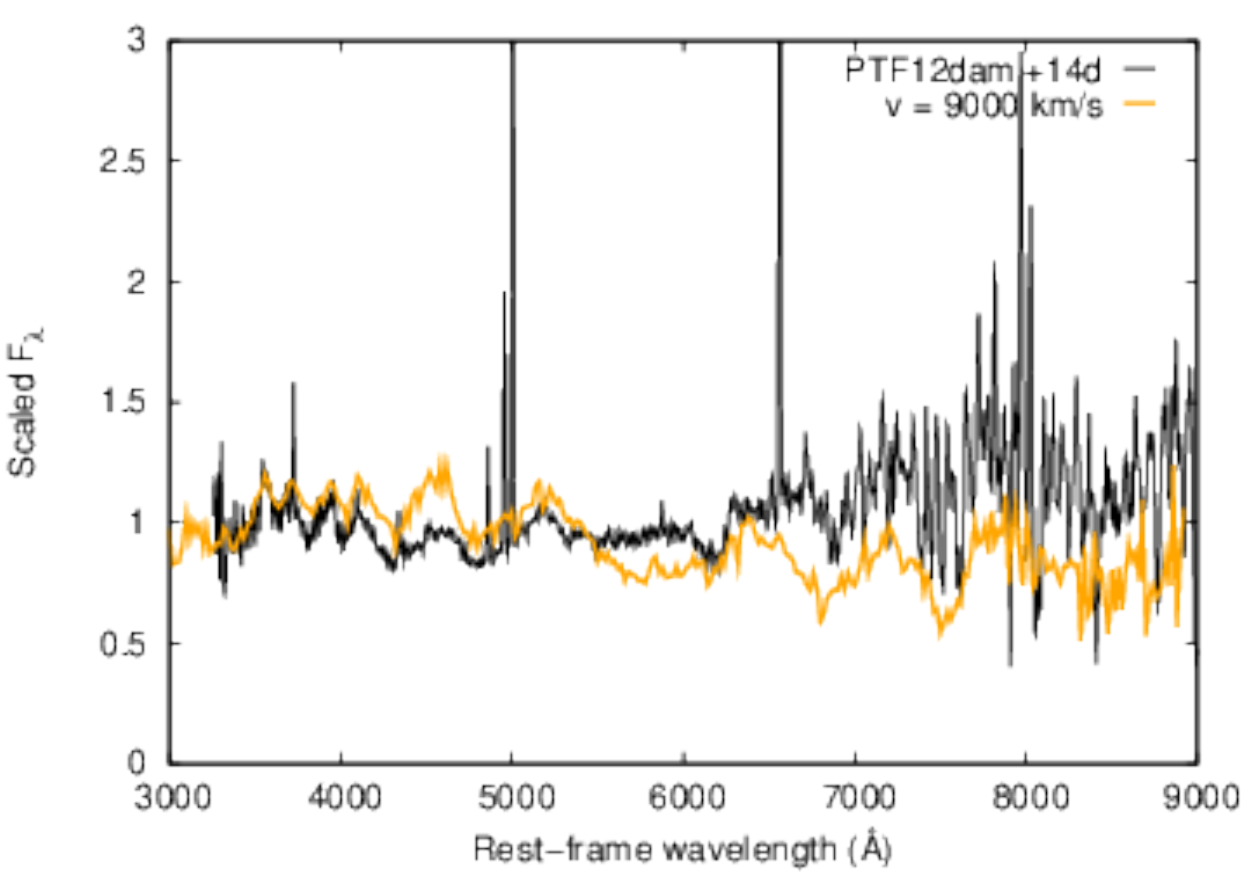}
\includegraphics[width=4cm]{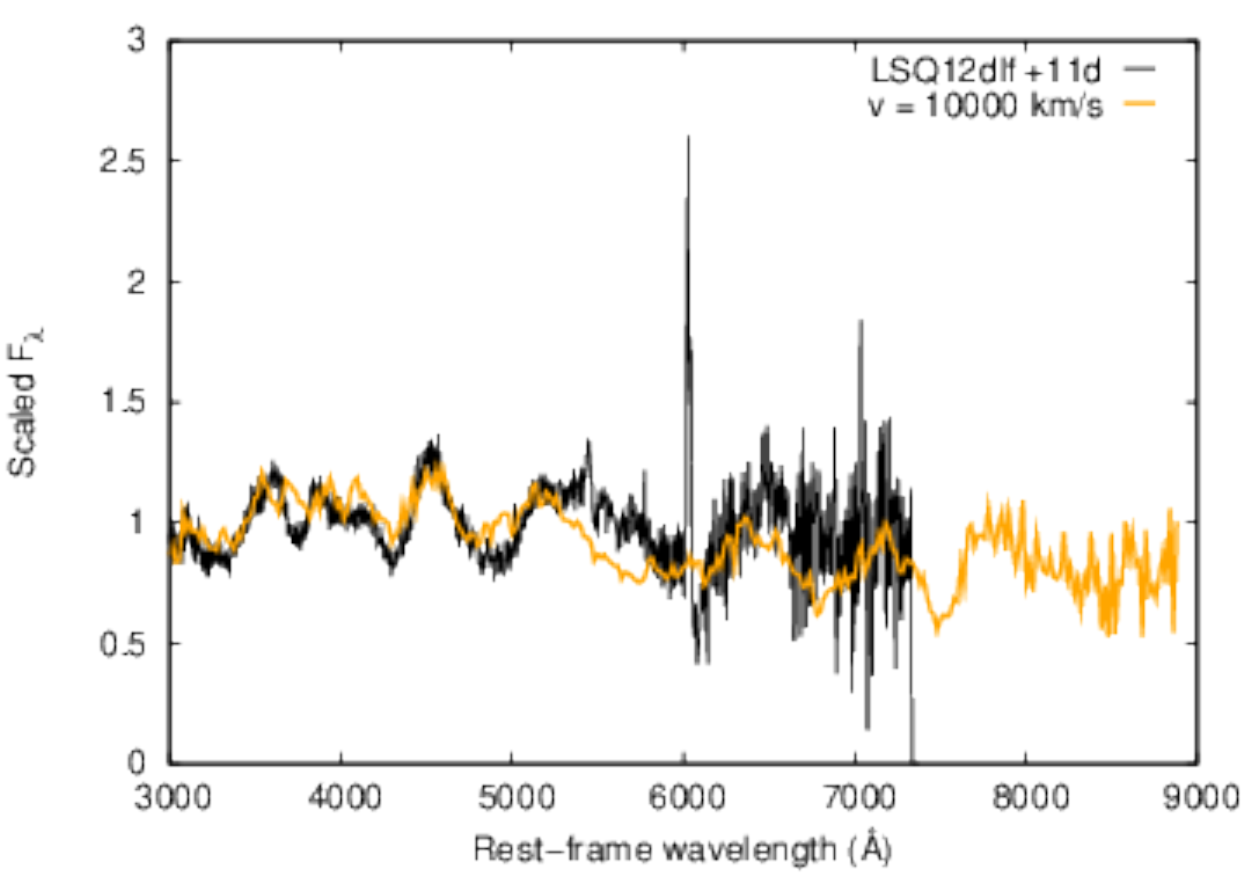}
\includegraphics[width=4cm]{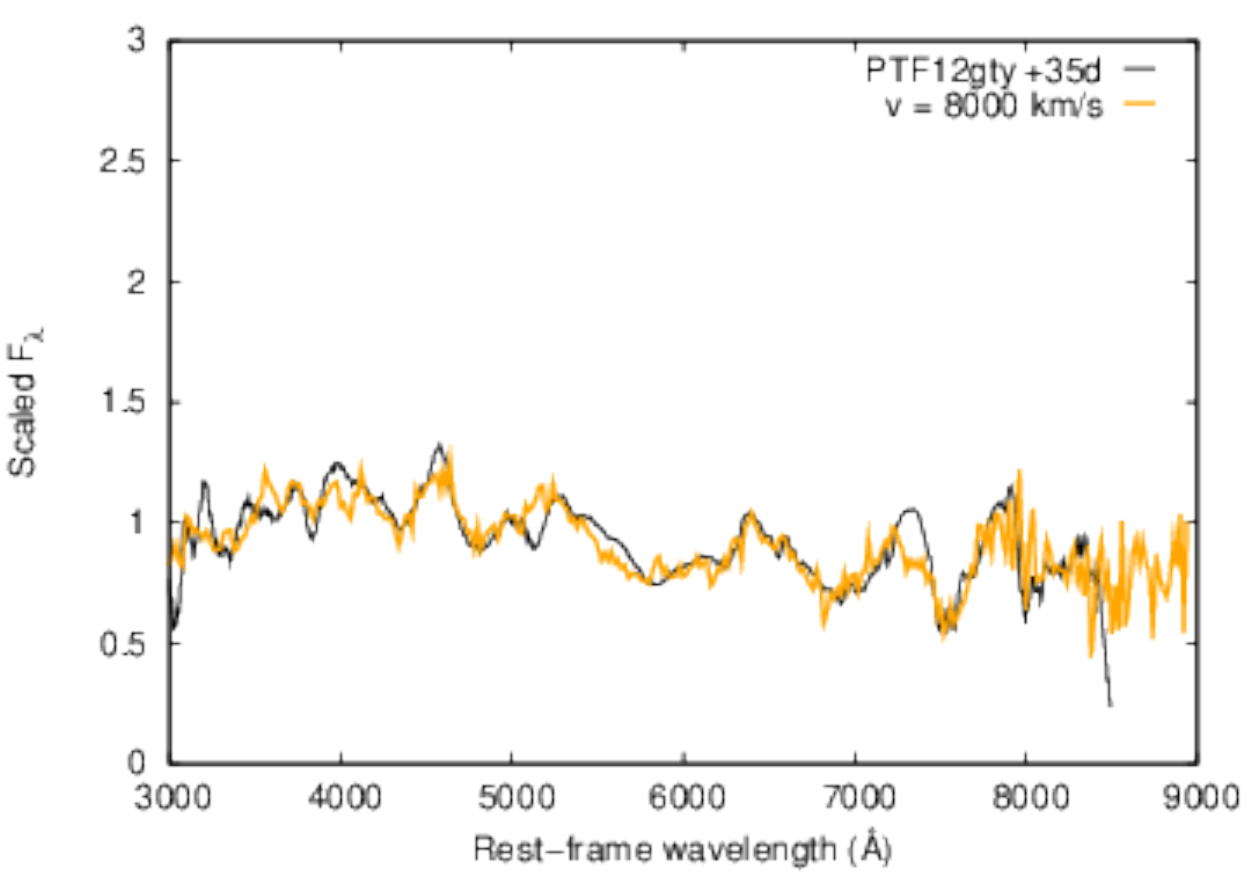}
\includegraphics[width=4cm]{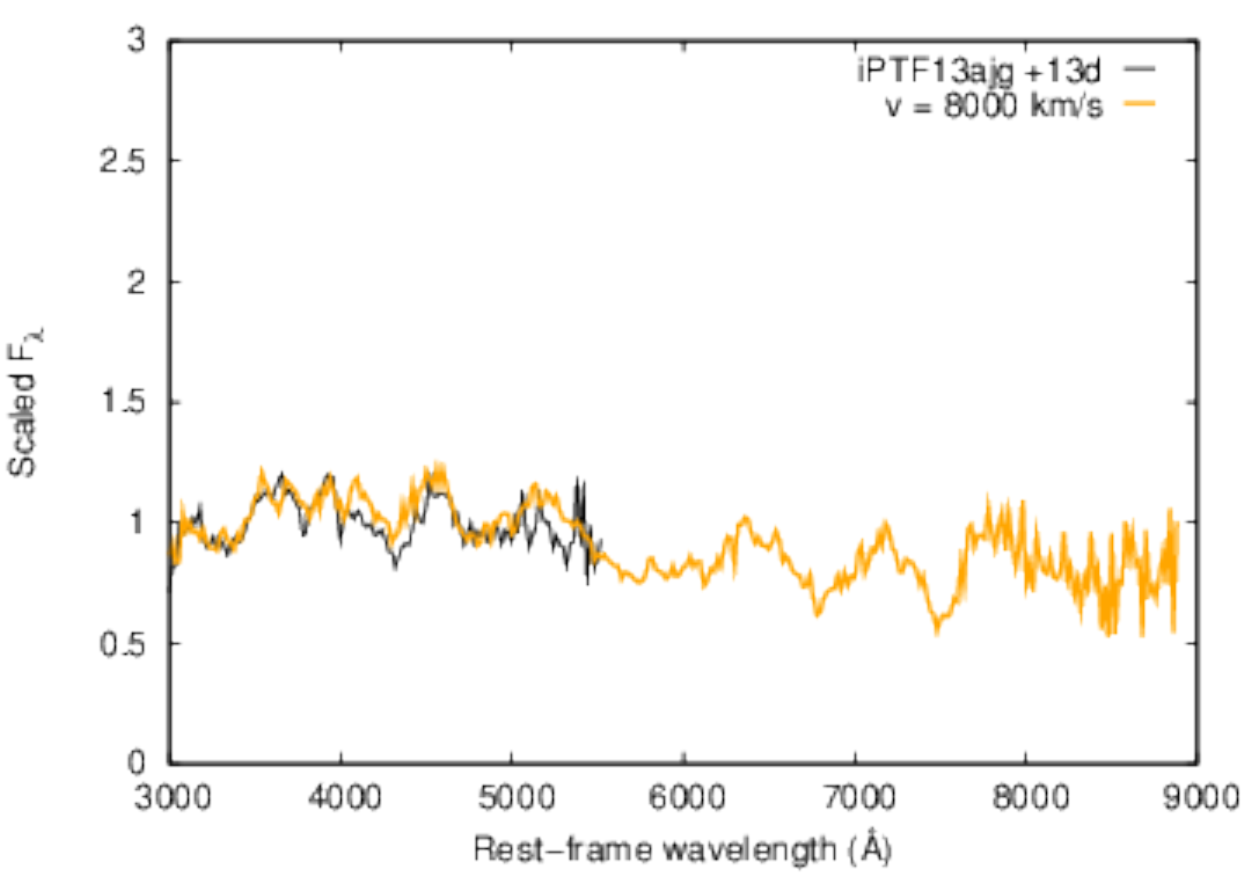}
\includegraphics[width=4cm]{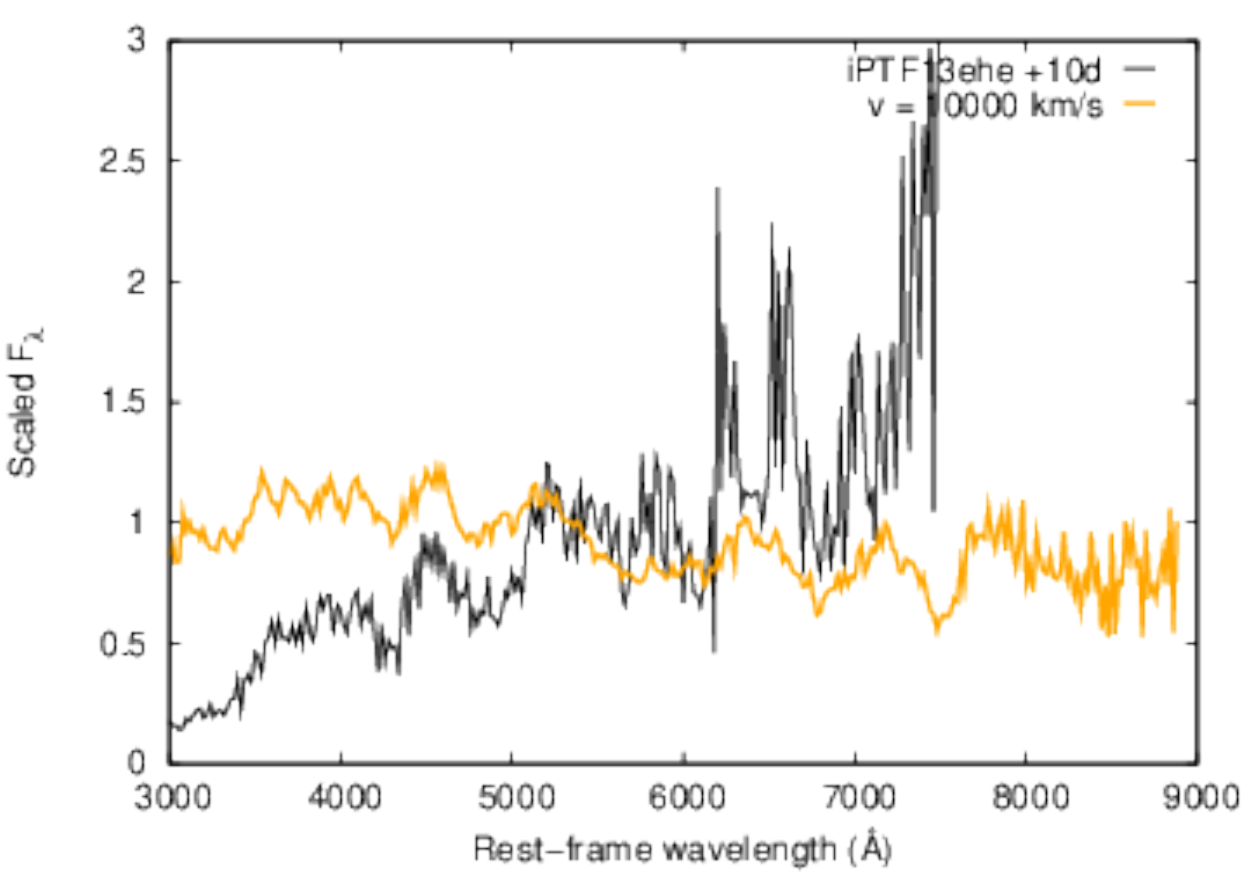}
\includegraphics[width=4cm]{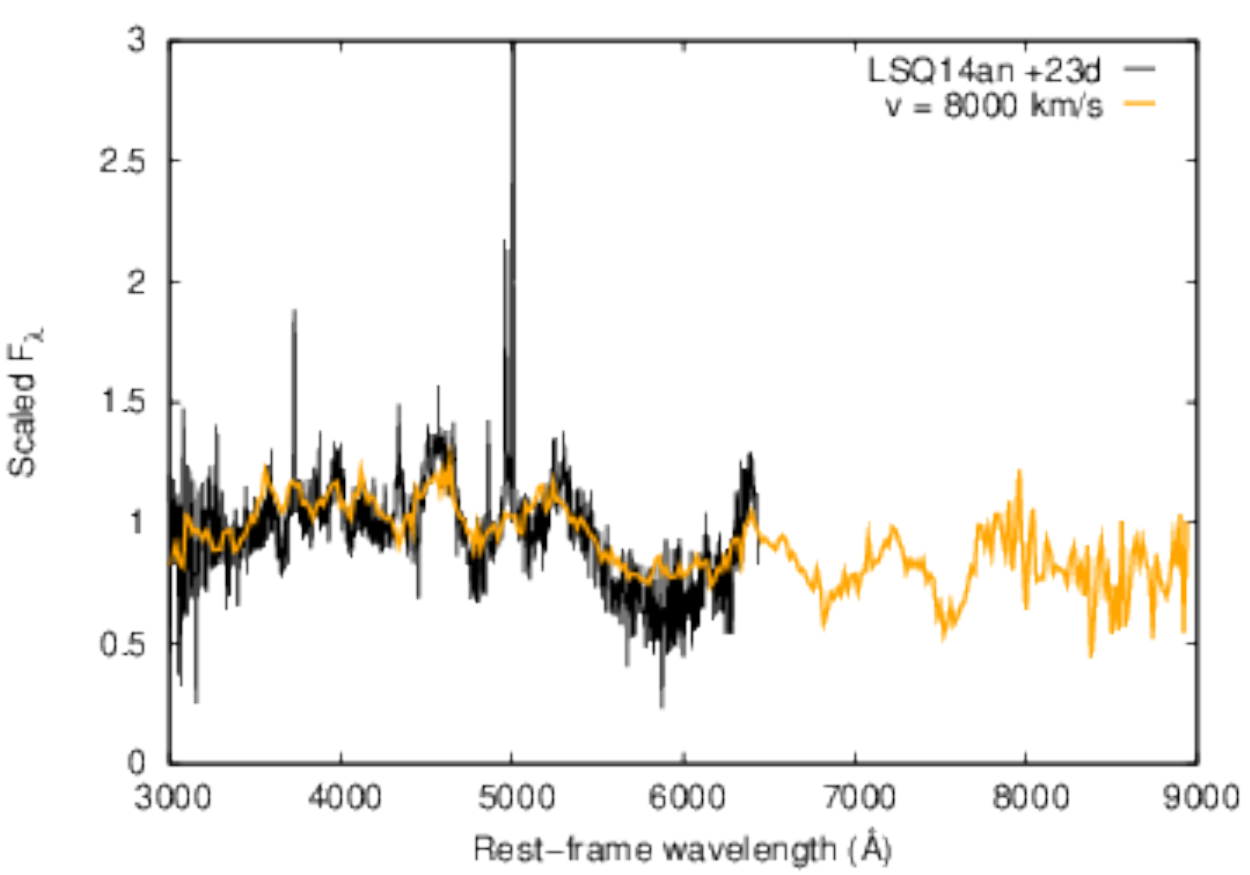}
\includegraphics[width=4cm]{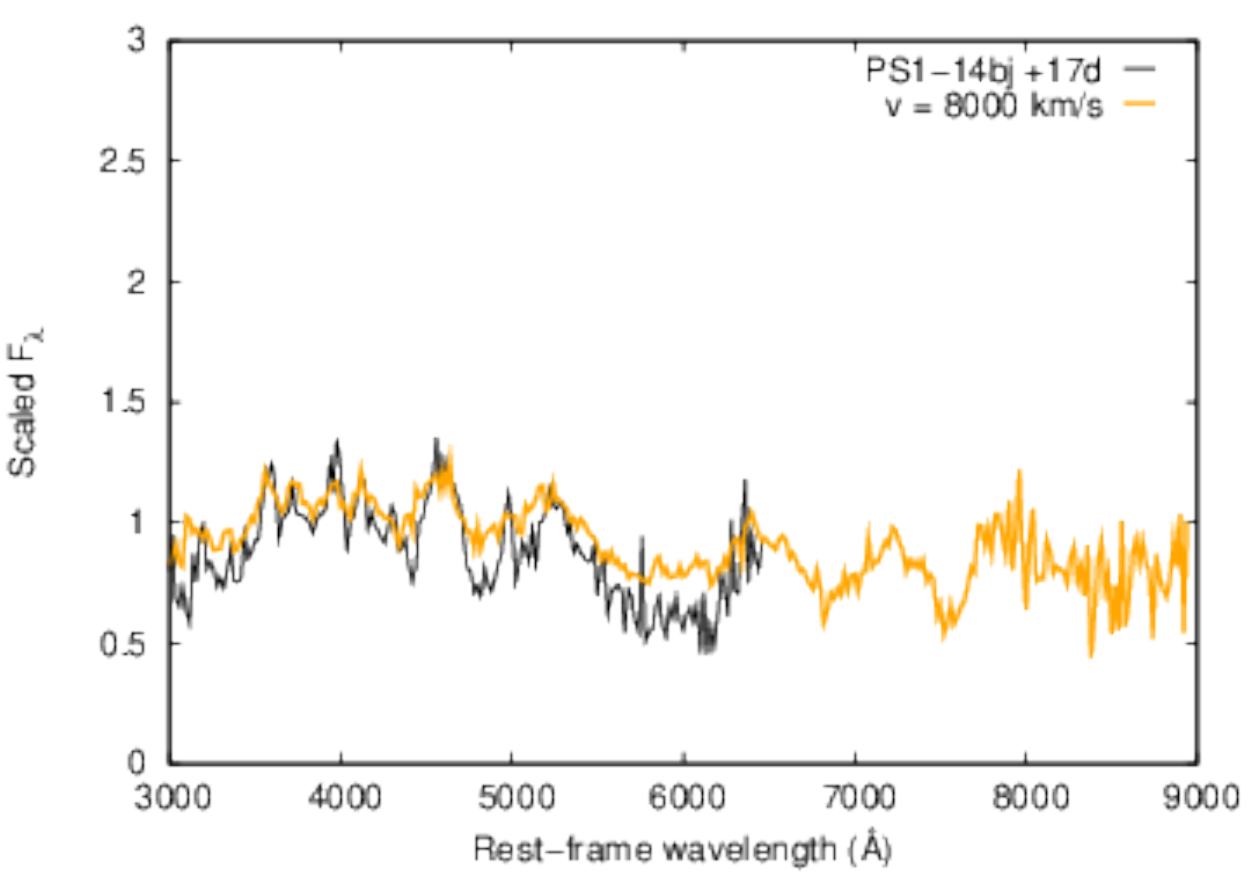}
\includegraphics[width=4cm]{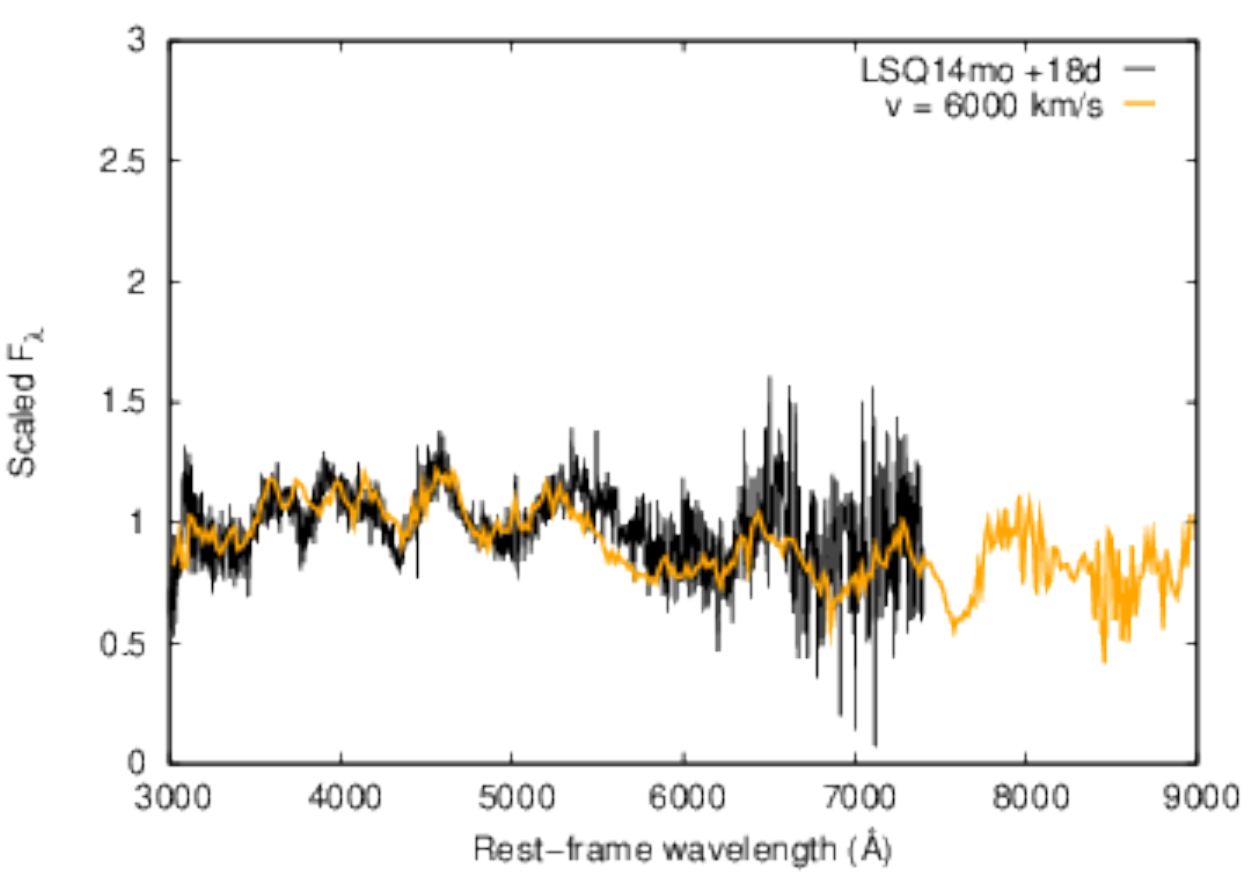}
\includegraphics[width=4cm]{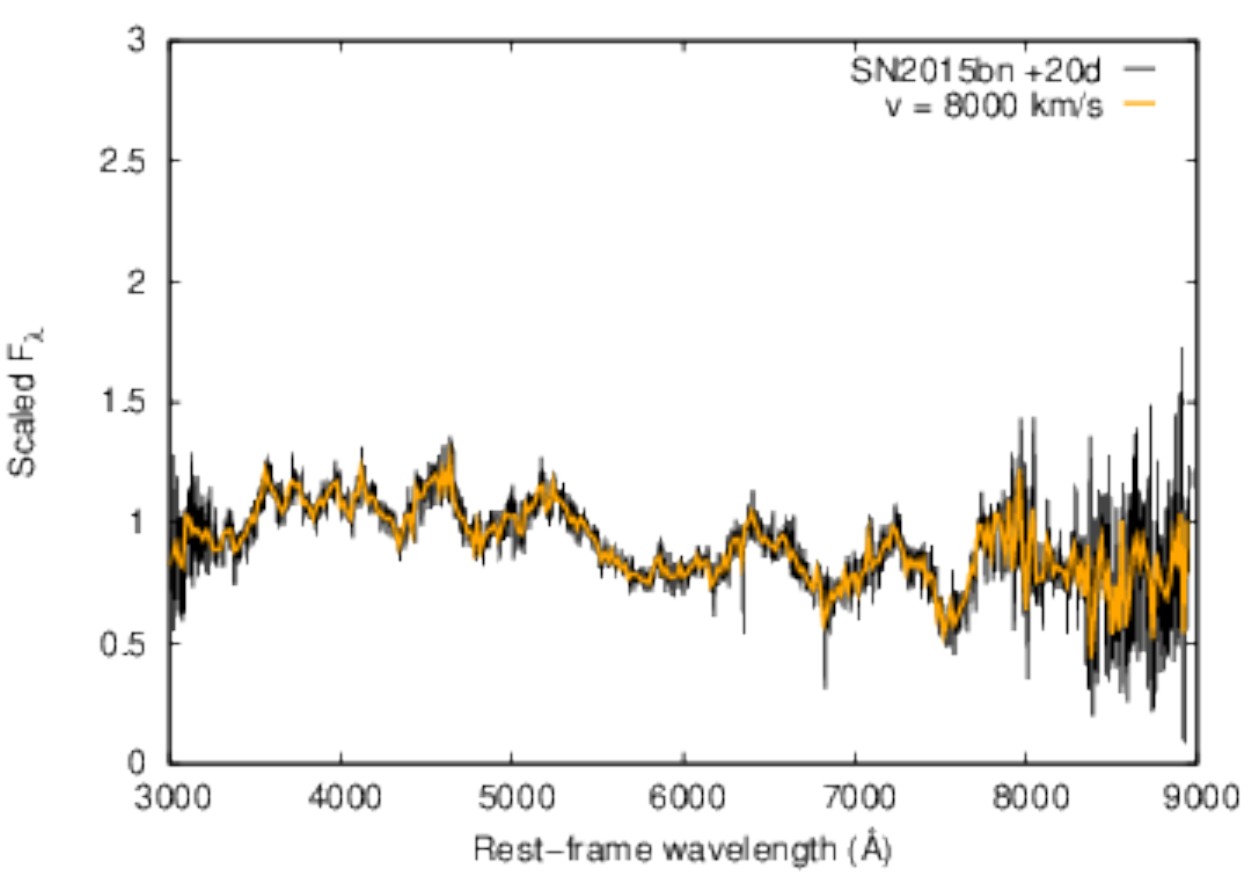}
\includegraphics[width=4cm]{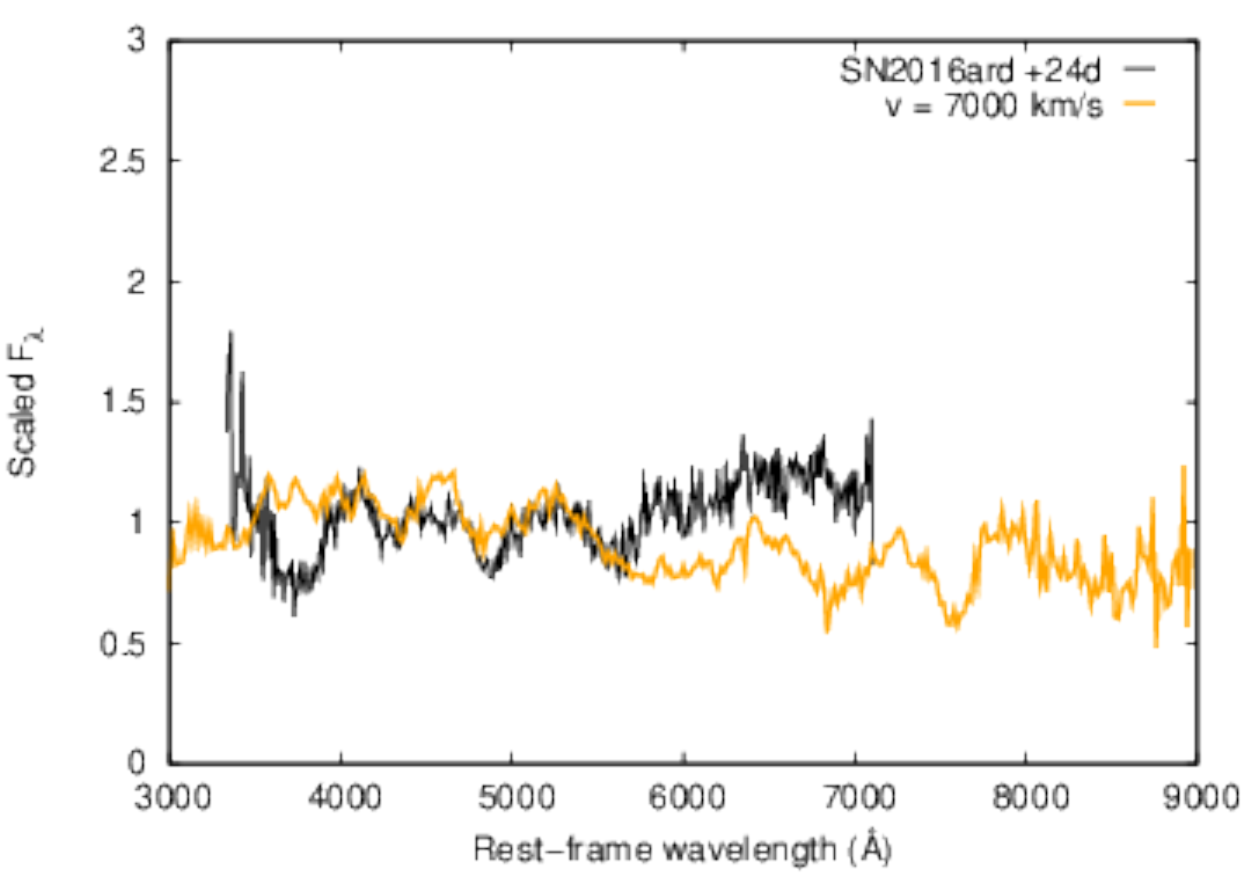}
\includegraphics[width=4cm]{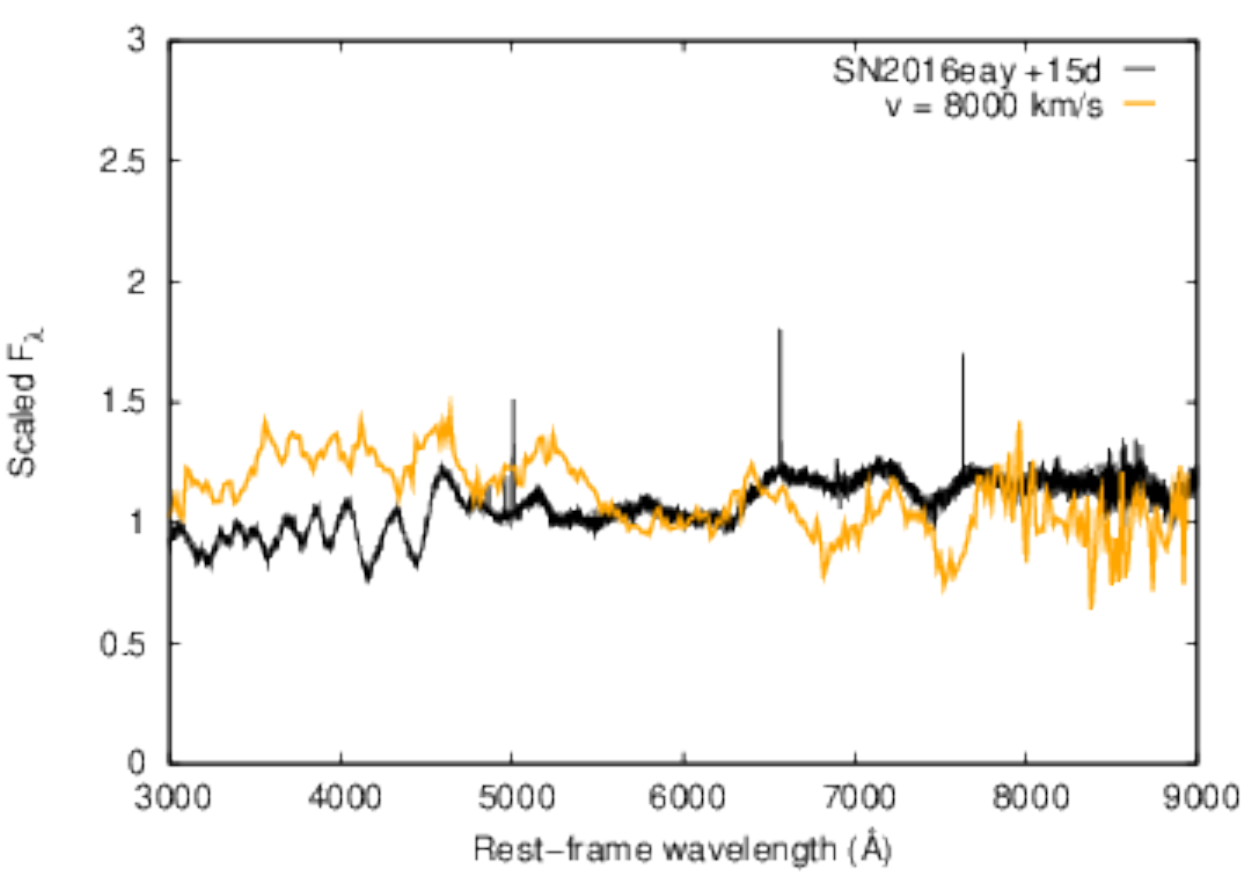}
\includegraphics[width=4cm]{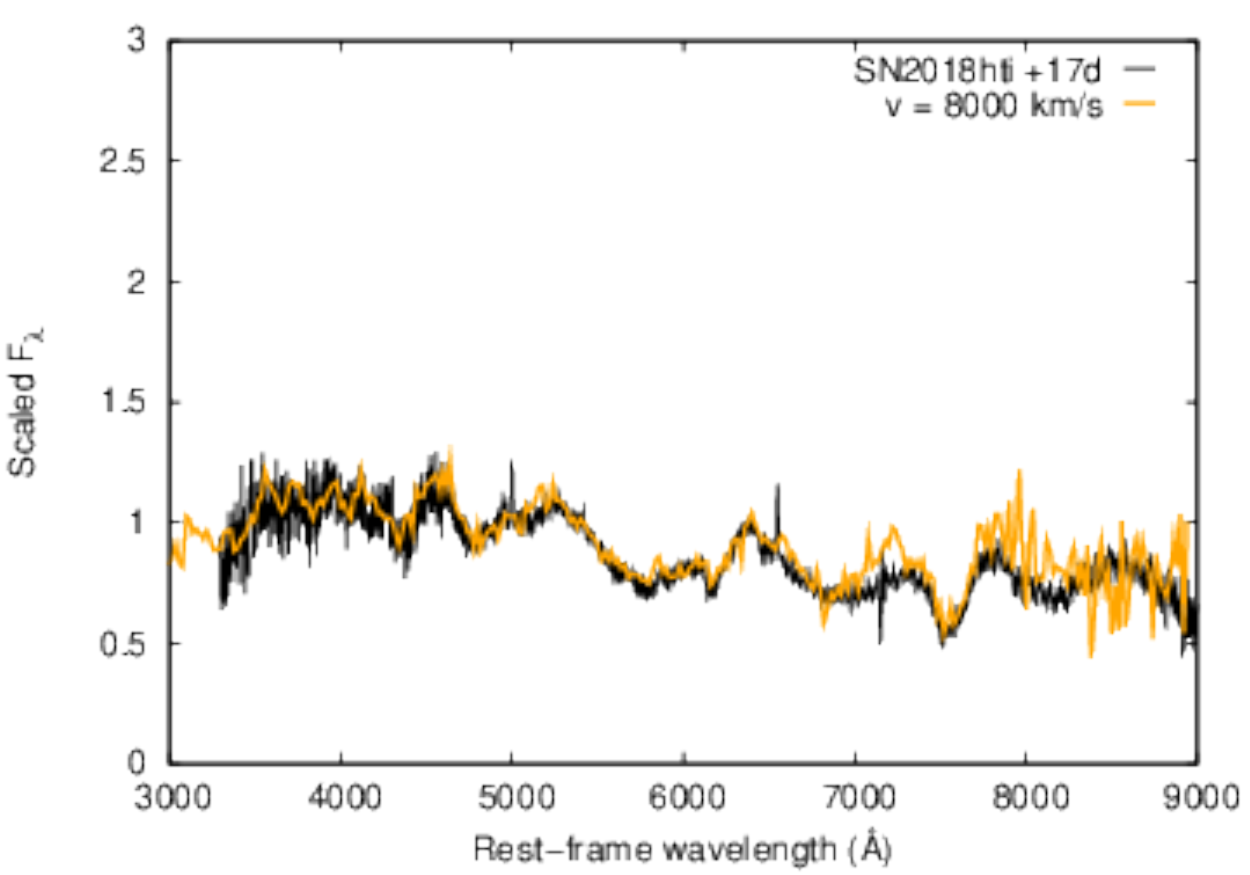}
\includegraphics[width=4cm]{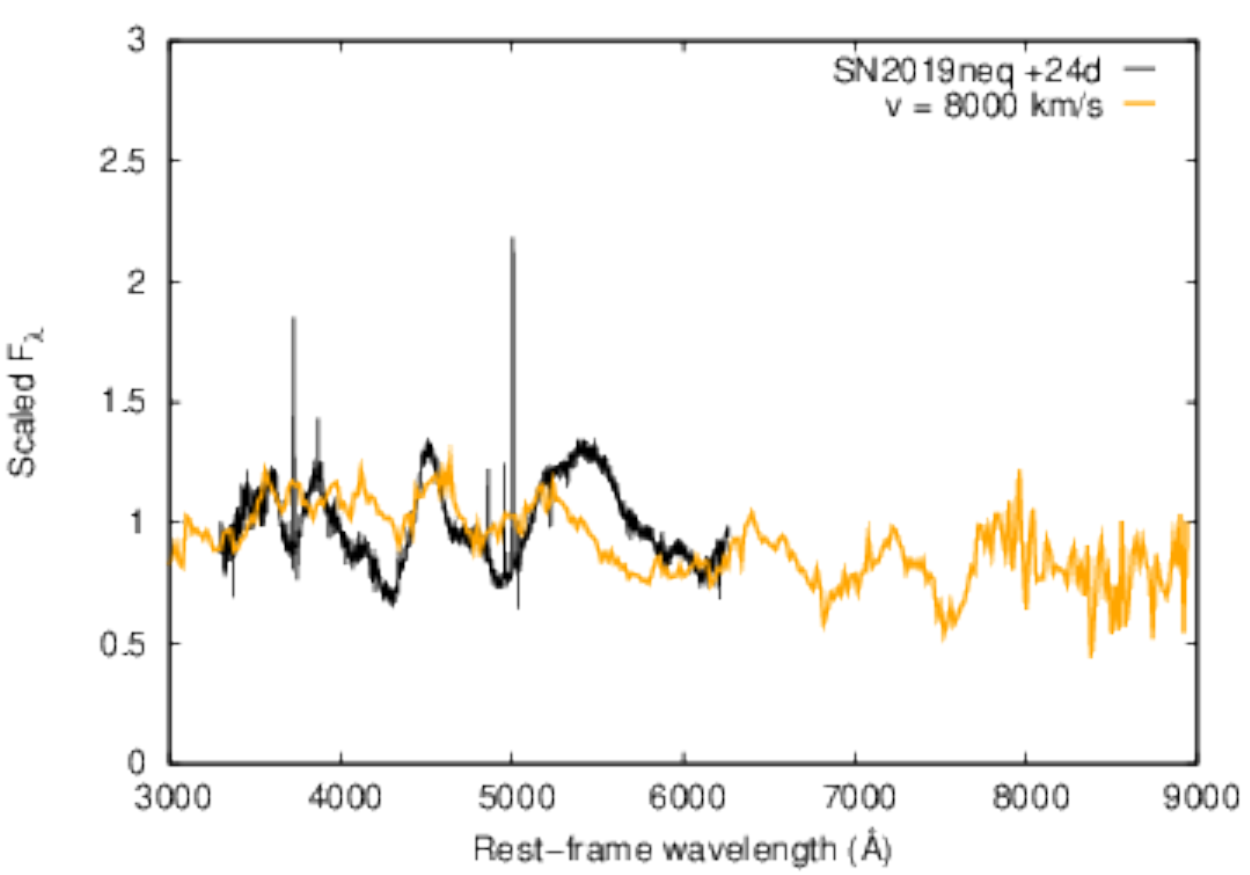}
\caption{The corrected and normalized "later post-maximum" phase spectra of the examined objects (black lines) plotted together with their best-fit template spectrum (orange lines) used in the cross correlation process. }
\label{fig:korr}
\end{figure*}

\begin{figure*}
\centering
\includegraphics[width=8cm]{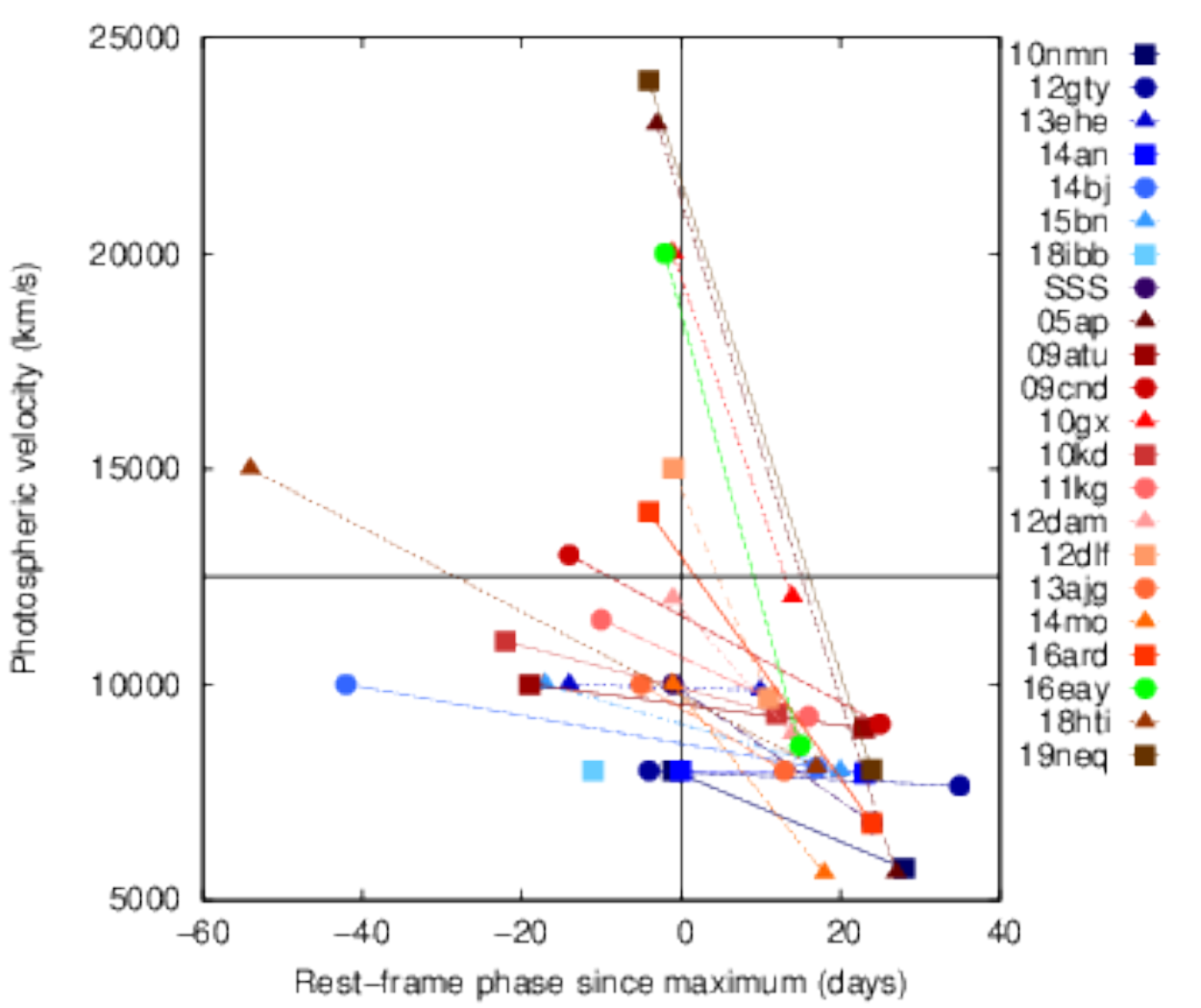}
\includegraphics[width=8cm]{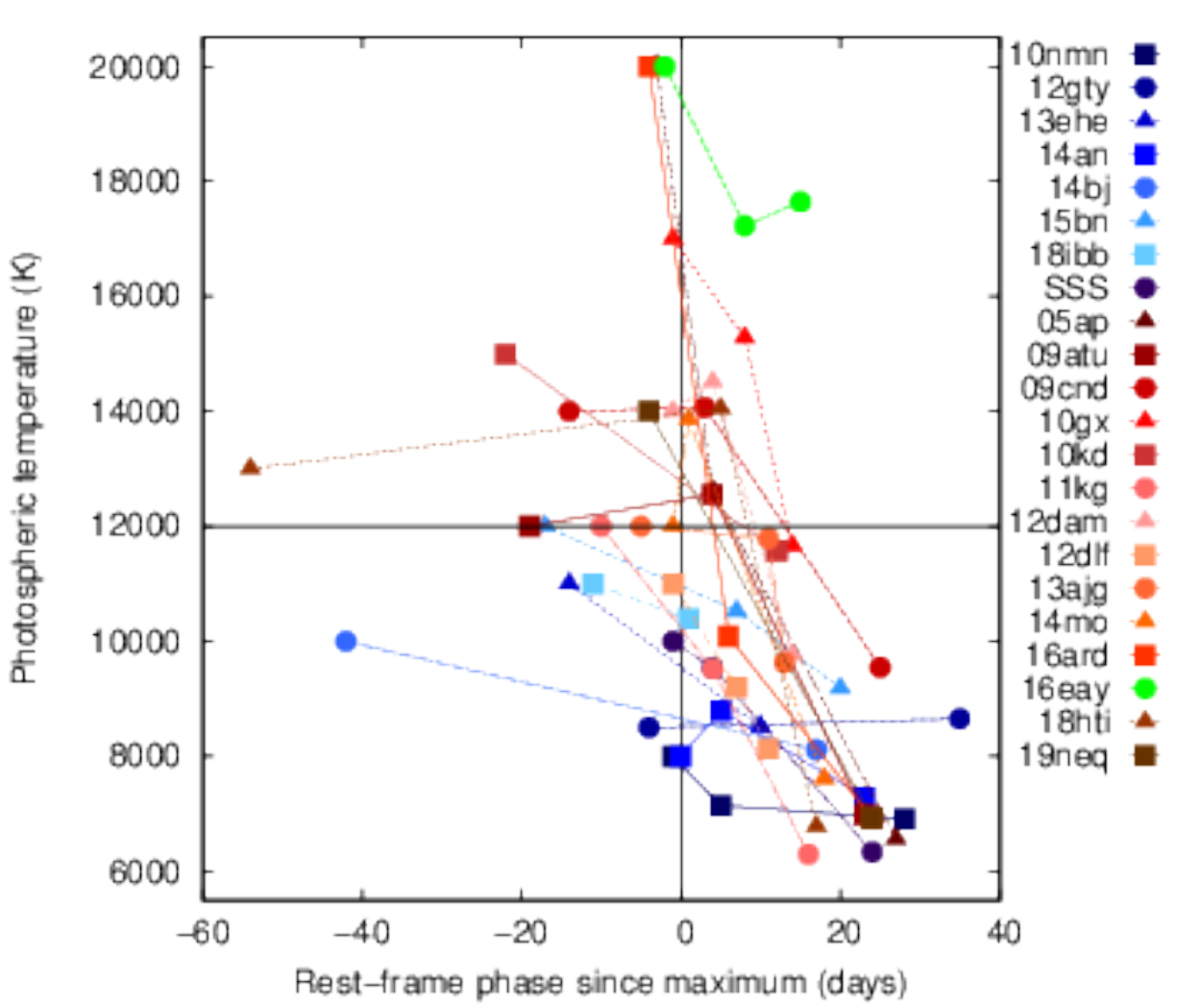}
\caption{Left panel: Photospheric velocity evolution of the SLSNe-I in the sample. Pre-maximum $v_{\rm phot}$ values were adopted from \citet{ktr21}, while "later post-maximum" photospheric velocities were estimated via cross correlation. Type W SLSNe-I are coded with red colors, while Type 15bn objects are shown with blue. As a peculiar object that shows the W-shaped feature in its spectra in the post-maximum epoch as well, SN~2016eay is marked with green color. For clarity, the epoch of the maximum light (black vertical line), and $v_{\rm phot}$ = 12000 km s$^{-1}$ (black horizontal line) is displayed as well. Right panel: Photospheric temperature evolution of the studied objects, using the same color coding as in the left panel. Here, the black horizontal line codes the $T_{\rm phot}$ = 12000 K value. It can be seen that despite the pre-maximum bimodality, Type W and Type 15bn SLSNe-I  show similar $v_{\rm phot}$ and $T_{\rm phot}$ values by the "later post-maximum" phases.}
\label{fig:grad}
\end{figure*}

\begin{figure}
\centering
\includegraphics[width=8cm]{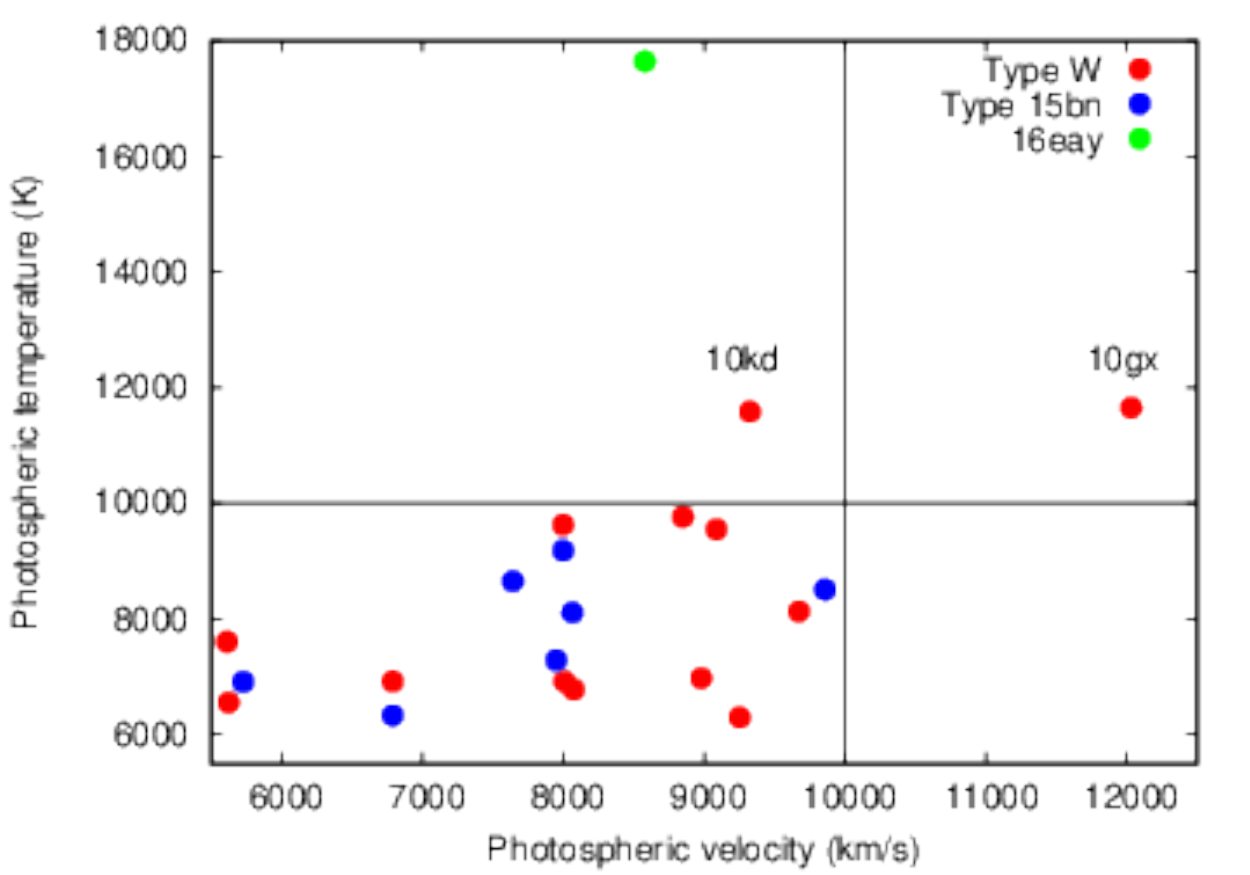}
\caption{Photospheric velocity vs photospheric temperature of the studied Type W (red) and Type 15bn (blue) SLSNe-I in the "later post-maximum" phase. Green color denote the outlier, SN~2016eay. 10000 K and 10000 km s$^{-1}$ are marked with black horizontal and vertical line, respectively. It is seen that the $T_{\rm phot}$ and $v_{\rm phot}$ of most objects decreased below 10000 K and 10000 km$^{-1}$, with the exception of SN~2010kd, SN~2010gx and SN~2016eay (all of them are Type W SLSNe-I).}
\label{fig:tph_vs_vph}
\end{figure}

Figure \ref{fig:templet} shows the template spectrum used during the cross-correlation. Here, the +20d spectrum of SN~2015bn ($v_{\rm phot}$ = 8000 km s$^{-1}$) is shifted in the velocity space in order to obtain templates having photospheric velocities between 4000 and 14000 km s$^{-1}$. In Figure \ref{fig:korr} the redshift- and extinction corrected, continuum normalized spectrum of the "later-post-maximum" phase of each studied object (black line) is plotted together with the template that has the most similar photospheric velocity value (orange line). It is seen that most spectra indeed look very similar to the +20d spectrum of SN~2015bn. The best-fit $v_{\rm phot}$ values obtained by this process are collected in Table \ref{tab:phases}, where the pre-maximum $v_{\rm phot}$ values adopted from \citet{ktr21} are present as well. 

The $v_{\rm phot}$ evolution of each object from the pre-maximum to the post-maximum epochs is plotted in the left panel of Figure \ref{fig:grad}. Type W SLSNe-I are shown with red colors, while blue colors denote to Type 15bn SLSNe-I. SN~2016eay, the only SLSN-I in the sample, where the W-shaped feature did not disappear until $\sim$+40d phase is highlighted with green. From this figure, the following phenomenon can be seen: Type 15bn SLSNe-I generally show lower $v_{\rm phot}$ compared to Type W SLSNe-I, which evolve slowly, or remains nearly constant by the "later post-maximum" phases. On the other hand, some Type W SLSNe-I show steep photospheric velocity gradient and very high photospheric velocity in the pre-maximum phases that declines to $\sim$10000 - 12000 km s$^{-1}$ by the post-maximum phases, while other Type W SLSNe-I have similar $v_{\rm phot}$ evolution to the Type 15bn SLSNe-I. It is seen that by the "later post-maximum" phase, the pre-maximum diversity/bimodality ceased to exist, and the $v_{\rm phot}$ of all studied SLSNe-I decreases below 12000 km s$^{-1}$. This implies that Type W and Type 15bn SLSNe-I cannot be distinguished by the post-maximum photospheric velocity.

This kind of homogenization remains true for the evolution of the photospheric temperatures as well. With the same color coding as in the left panel, the right panel of Figure \ref{fig:grad} displays the $T_{\rm phot}$ evolution of the studied SLSNe-I. Pre-maximum $T_{\rm phot}$ values were adopted from \citet{ktr22}, while the "early post-maximum", and "later post-maximum" photospheric temperatures were calculated by fitting a blackbody curve  to the spectrum of each object during the continuum normalization process. The $T_{\rm phot}$ values in the pre-maximum, the early, and the later post-maximum phases can be found in Table \ref{tab:phases}. From the right panel of Figure \ref{fig:grad}, one of the main conclusions of \citet{ktr22} can be seen as well: in the pre-maximum phases, Type W SLSNe-I tend to show $T_{\rm phot} >$  12000 K, while Type 15bn SLSNe-I have $T_{\rm phot} <$  12000 K. Some kind of bimodality is present in the "early post-maximum" phases, where the studied SLSNe-I show a wide range of photospheric temperatures (see Table \ref{tab:phases} as well). However, by the "later post-maximum" phases, the photospheric temperature of most examined objects become similar, and decrease to $<$ 10000 K. This is consistent with the observed spectra becoming homogeneous by the "later post-maximum" phases, and shows that the differences between Type W and Type 15bn SLSNe-I seem to disappear. SN~2016eay is an outlier in this respect, as the photospheric temperature of the object remains extremely high ($\sim$18000 K) by +25 days after maximum, and accordingly, the W-shaped feature remains dominant in the spectrum formation. 

The photospheric velocity vs photospheric temperature comparison of the "later post-maximum phase" can be found in Figure \ref{fig:tph_vs_vph}. Type W SLSNe-I are plotted with red, while Type 15bn SLSNe-I are marked with blue color. The point referring to SN~2016eay is green. It can be seen, again that the sample becomes mostly homogeneous by the examined phase, so both the $v_{\rm phot}$ and the $T_{\rm phot}$ values decreased below 10000 km s$^{-1}$ and K respectively. Besides SN~2016eay, SN~2010kd and SN~2010gx (both Type W SLSNe-I) seem to be slightly different from the other members of the sample. While the temperature of these two SLSNe-I remain 12000 K by the "later post-maximum" phases, SN~2010gx shows larger $v_{\rm phot}$ compared to the other examined SLSNe-I. This suggests that maybe some of the Type W SLSNe-I remain different from the Type 15bn objects in the post-maximum phases, but most of them become similar to them. 

As a conclusion, we recognize that apart from SN~2016eay, both the photospheric velocities and temperatures of Type W and Type 15bn SLSNe-I became mostly homogeneous by $\sim$+30 days after the maximum.

\subsection{Pseudo-equivalent width calculations}

Inferring the pseudo-equivalent widths (pEW) of particular parts/features of the spectra may reveal additional information on the physical differences between Type W and Type 15bn SLSNe-I in the post-maximum phases. Therefore, we disposed three characteristic part of the examined spectra, and derived pseudo-equivalent widths via direct integration in the "early post maximum", the "later post-maximum" and the "pseudo-nebular" phases using the {\tt splot} task in the {\tt onedspec} package of IRAF \footnote{IRAF is distributed by the National Optical Astronomy
Observatories, which are operated by the Association of Universities for Research in Astronomy, Inc., under cooperative agreement with the National Science Foundation. http://iraf.noao.edu} (Image Reduction and Analysis Facility).

\begin{figure*}
\centering
\includegraphics[width=12cm]{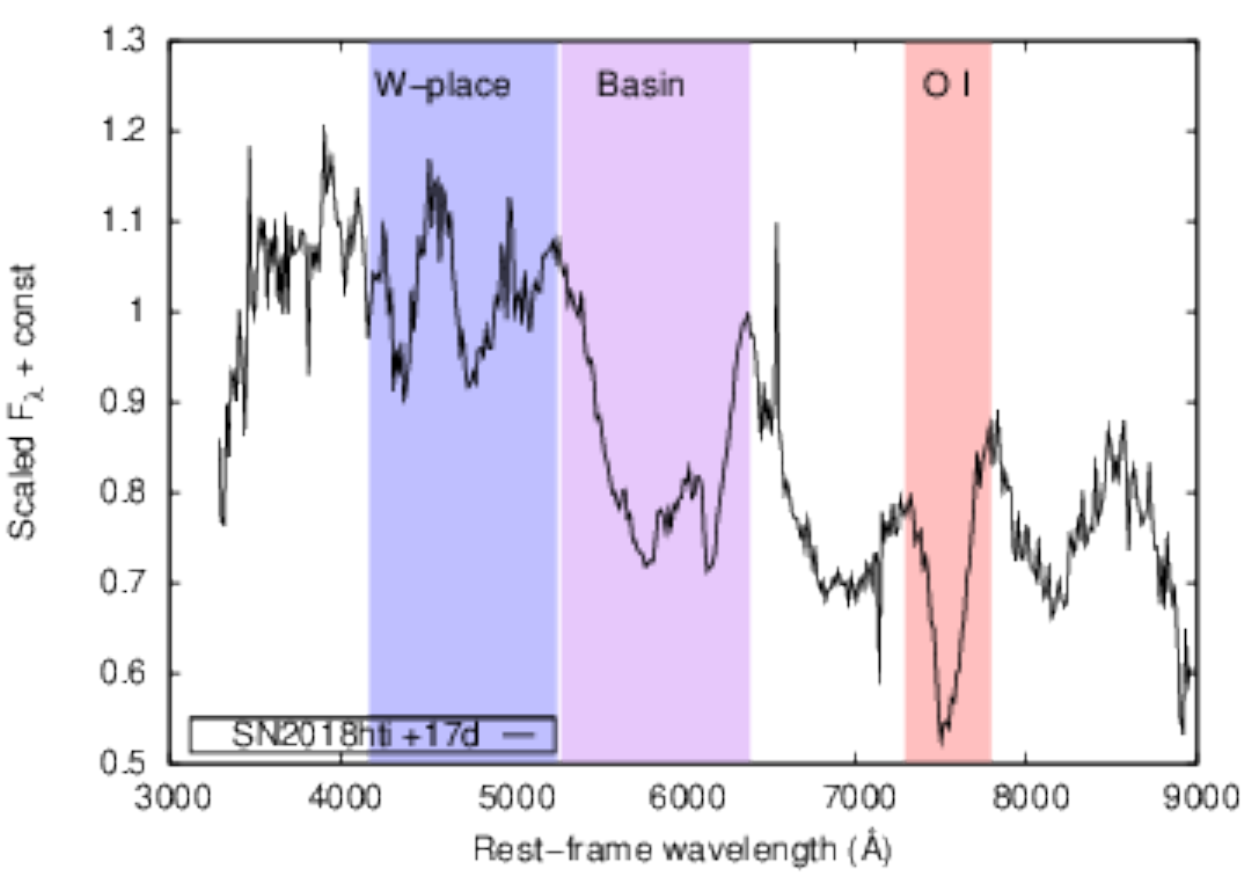}
\caption{The +17d phase spectrum of SN~2018hti. The 3 regions used for pEW calculations are shaded with different colors: the wavelength range between 4166 and 5266 \AA, where the W-shaped O II blend before the maximum is supposedly present in case of Type W object is marked with blue, the basin-shaped feature present in all examined SLSNe-I in the post-maximum phases between 5290 and 6390 \AA is plotted with purple, while red color denote the place of the commonly present O I line between 7295 and 7797 \AA. }
\label{fig:18hti_ew}
\end{figure*}

The three characteristic regions of the spectrum of SN~2018hti (+17d phase) are shaded with different colors in Figure \ref{fig:18hti_ew}. The first region  (colored with blue, ranging between 4166 and 5266 \AA)  denote the wavelengths, where the W-shaped feature should be present in the case of Type W SLSNe-I during the pre-maximum phases. Therefore, in the following, we name it "W-place". The pEW of this region was also calculated in Type 15bn SLSNe-I for comparison in spite of the missing W-shaped feature. The second examined blend is the region between 5290 and 6390 \AA\ (shaded with purple in Figure \ref{fig:18hti_ew}), and it is strongly present in both Type W and Type 15bn SLSNe-I in the post-maximum phases (see e.g. the right panel of Figure \ref{fig:earlypost}. In the following, we are  going to refer to this feature as the "basin". The third region is the place of the O I $\lambda$7775 line (7295 -- 7797 \AA, shaded with red), which is present in case of most SLSNe-I both before and after the maximum. 

\begin{figure*}
\centering
\includegraphics[width=8cm]{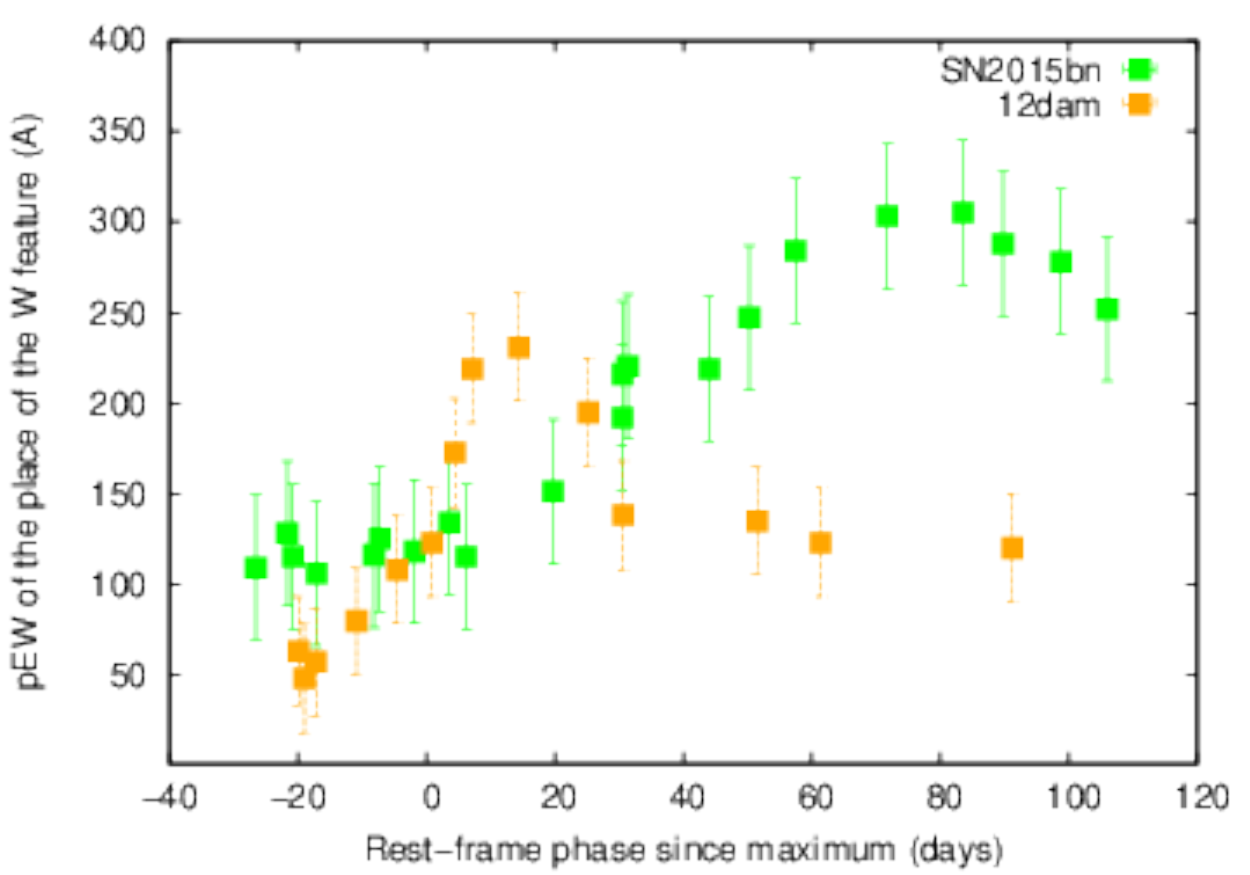}
\includegraphics[width=8cm]{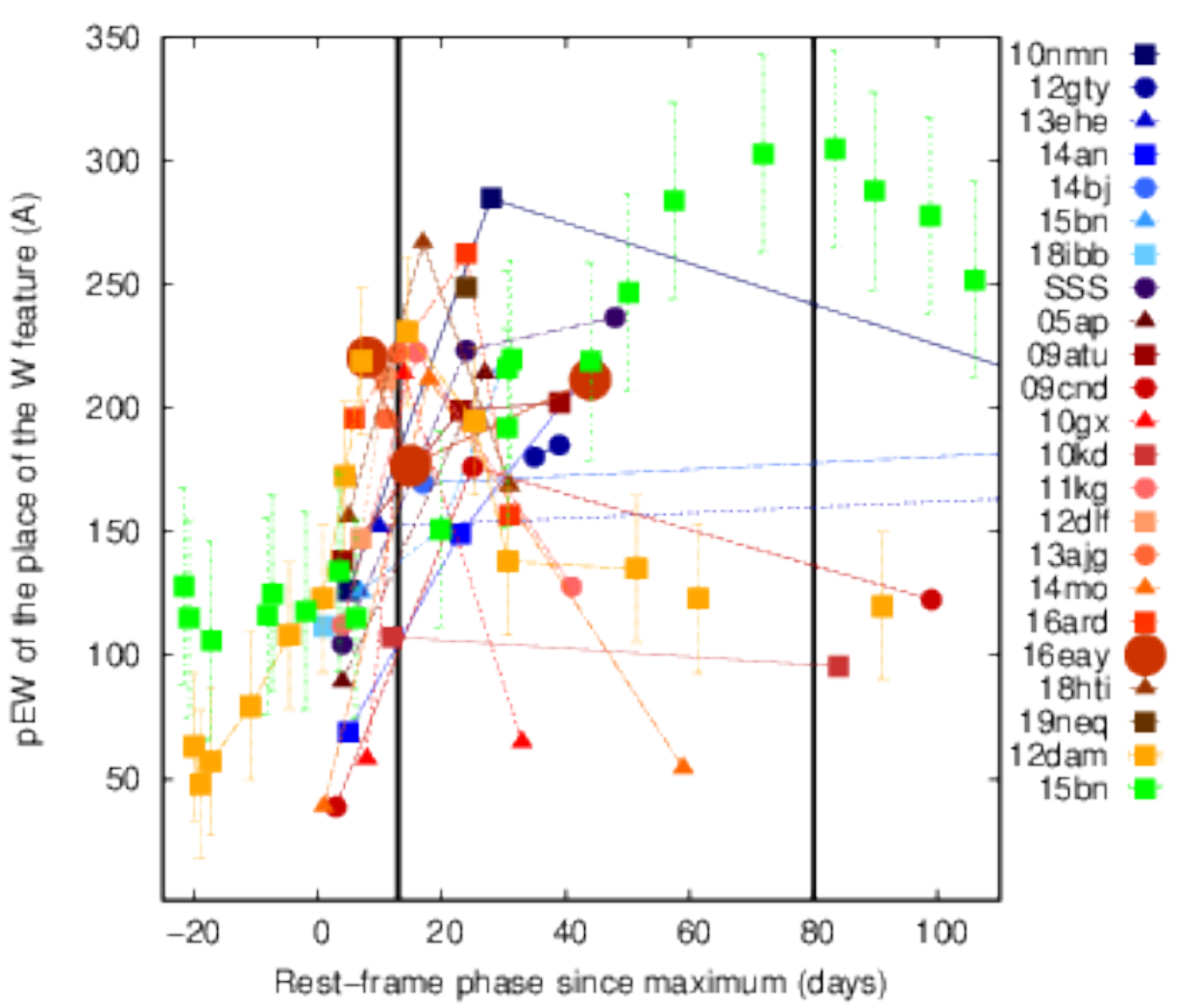}
\caption{Left panel: pEW calculations of the "W-place" feature for all available spectra of SN~2015bn (green) and PTF12dam (orange). Right panel: evolution of the pEW of all studied SLSNe-I in the "early post-maximum", the "later post-maximum" and the pseudo-nebular phases. Type W SLSNe-I are coded with red, while blue colors denote to the Type 15bn objects. The symbol of SN~2016eay, the outlier is enlarged. for clarity. The vertical black lines shows the supposed maximum place of the pEW of the W-place feature for Type W SLSNe-I ($\sim$+13d phase), and Type 15bn SLSNe-I ($\sim$ +80d phase).}
\label{fig:15bn_ew}
\end{figure*}

\begin{figure*}
\centering
\includegraphics[width=8cm]{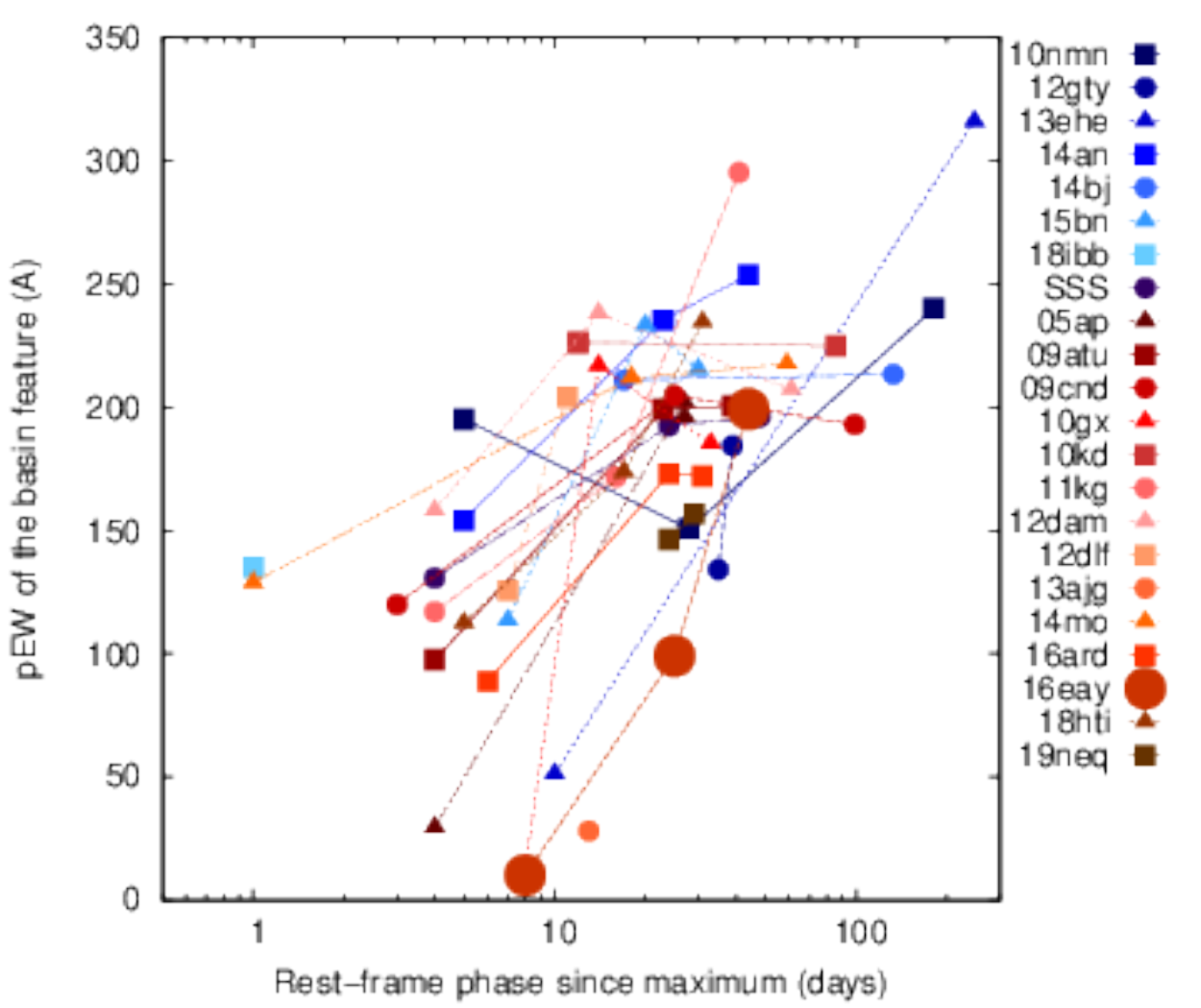}
\includegraphics[width=8cm]{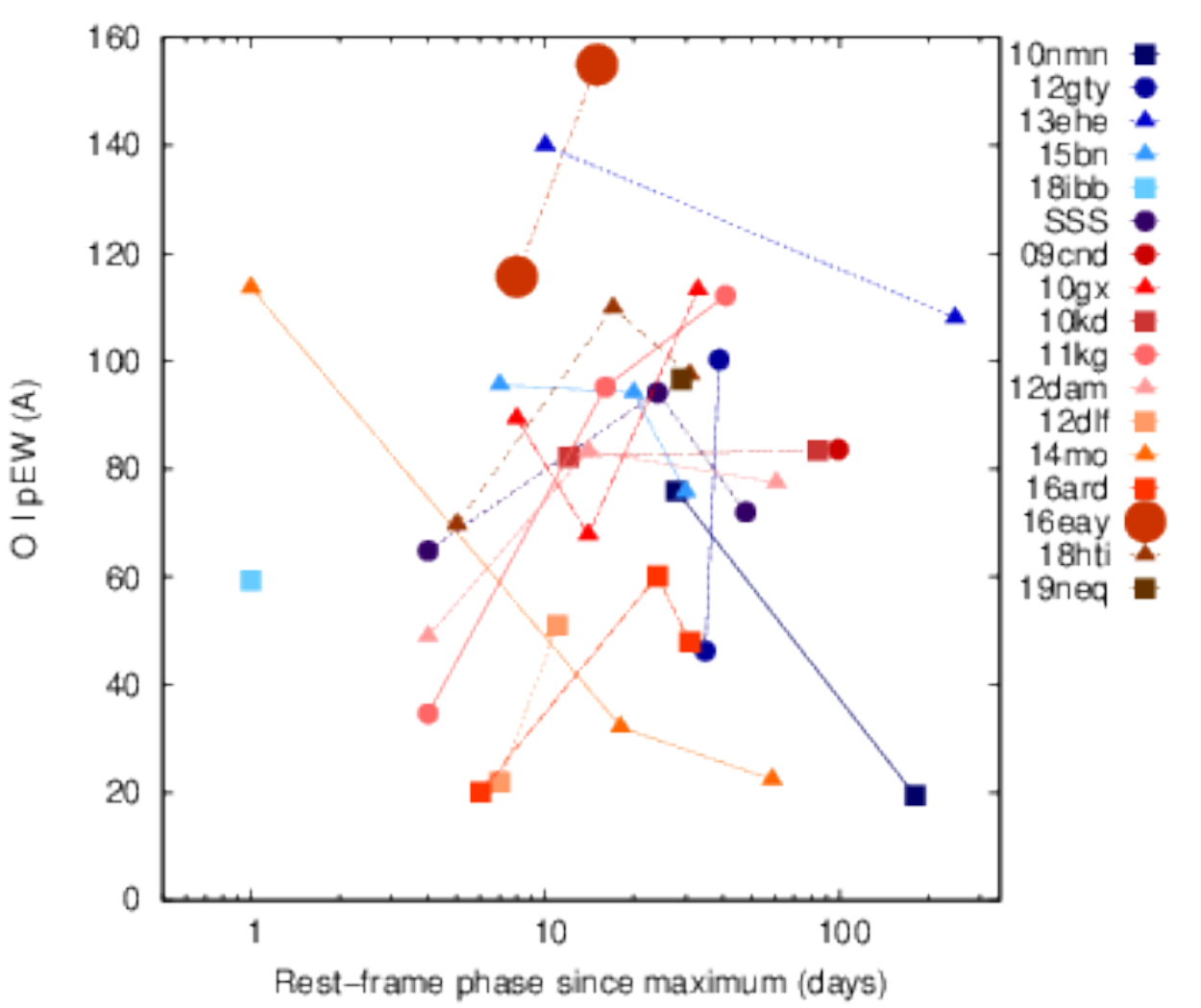}
\caption{Left panel: pEW evolution of the "basin" feature for all studied SLSNe-I. Right panel: pEW evolution of the O I $\lambda$7775 line for all studied objects. The color coding is the same as in Figure \ref{fig:15bn_ew}.}
\label{fig:ew}
\end{figure*}

For most studied SLSNe-I, only a few post-maximum spectra were available, therefore examining the evolution of the pEW of the mentioned regions has to be treated with caution. For unambiguous conclusions, the examination of high-cadence data from the pre-maximum to the nebular phases is planned in the future, but it is beyond the scope of the present paper. Here, the extensive pEW calculations from the earliest to the latest available photospheric phase spectra is presented for two objects having a lot of data: SN~2015bn (Type 15bn SLSN-I), and PTF12dam (Type W SLSN-I) for the "W-place" region. Since this wavelength-range is the most different in case of the Type W and the Type 15bn groups,  differences in the pEWs may be another evidence for the bimodality between these two groups. The left panel of Figure \ref{fig:15bn_ew} shows the pEW evolution of the "W place" region for SN~2015bn (green) and PTF12dam (orange) between $\sim$ -35d and $\sim$ + 110d phase. It is seen that the pEW of the 2 objects evolves differently: the pEW of the W-shaped feature shows a maximum at $\sim$ +10d phase in case of the Type W objects, while it seems to be nearly constant in the pre-maximum phases, and increase by $\sim$ +80d for the Type 15bn SLSN-I. The right panel of Figure \ref{fig:15bn_ew} shows the evolution of the "W-place" feature for all studied objects for the "early post-maximum", the "later post-maximum" and the pseudo-nebular phases, plotted together with the values presented for SN~2015bn and PTF12dam in the left panel. Type W SLSNe-I are coded with red colors, while blue colors denote to Type 15bn SLSNe-I. The symbol of SN~2016eay is enlarged in order to make it more visible. The same trend as in the left panel can be detected for the Type W and the Type 15bn SLSNe-I, but it remains true that more data are needed to prove unambiguously that the two sub-groups are different in this respect.

Figure \ref{fig:ew} shows the evolution of the "basin" feature (left panel) and the O I line (right panel) of the studied SLSNe-I in the 3 examined epochs using the same color- and symbol coding as in the right panel of Figure \ref{fig:15bn_ew}. It is seen that the "basin" feature seems to get stronger with time in case of both Type W and Type 15bn SLSNe-I, so the region of the spectrum between $\sim$5000 and 6000 \AA\ evolves similarly in case of the two groups. On the contrary, no trend can be seen in the pEW-evolution of the O I line for either groups.

\subsection{Host galaxy properties}\label{sec:host}

In \citet{ktr22}, it was found that Type W and Type 15bn SLSNe-I may be different in the evolution of the early light curve and the geometry of the ejecta: some of the Type W SLSNe-I showed light curve undulations in the pre-maximum phases, while the rising light curves of Type 15bn SLSNe-I showed a tendency for smooth evolution. On the other hand, polarimetrically the Type W SLSNe-I that have data were concluded to be spherically symmetric, while some of the Type 15bn objects showed increasing polarization, suggesting distinct progenitor mechanisms for the two subgroups.  This raises the following question: are the explosion environment of the two groups different as well?

In order to search for possible answers to this question, data from the literature were collected for the objects in our sample. From the examined 27 SLSNe-I, 13 had public data regarding the properties of the host galaxy (e.g. the absolute brightness, the star formation rate, the stellar mass and the morphology), from which objects, 8 belong to the Type W group, while 5 were classified as Type 15bn. Table \ref{tab:host} shows the host galaxy properties of these SLSNe-I, together with their references. It can be read out from the table that most objects appear in faint or compact dwarf galaxies, consistently with other SLSNe-I. One counterexample, PTF10nmn exploded in an irregular galaxy.

\begin{figure*}
\centering
\includegraphics[width=6cm]{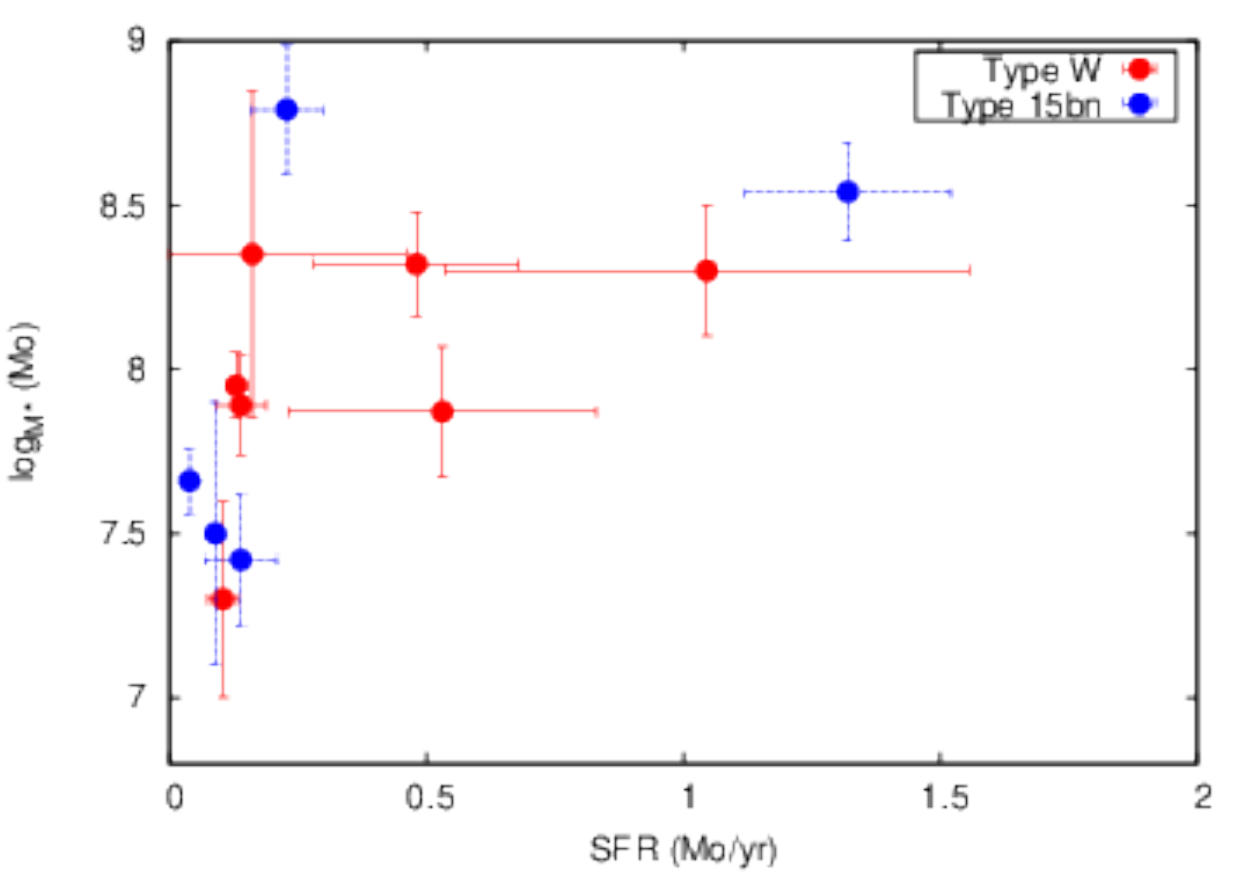}
\includegraphics[width=6cm]{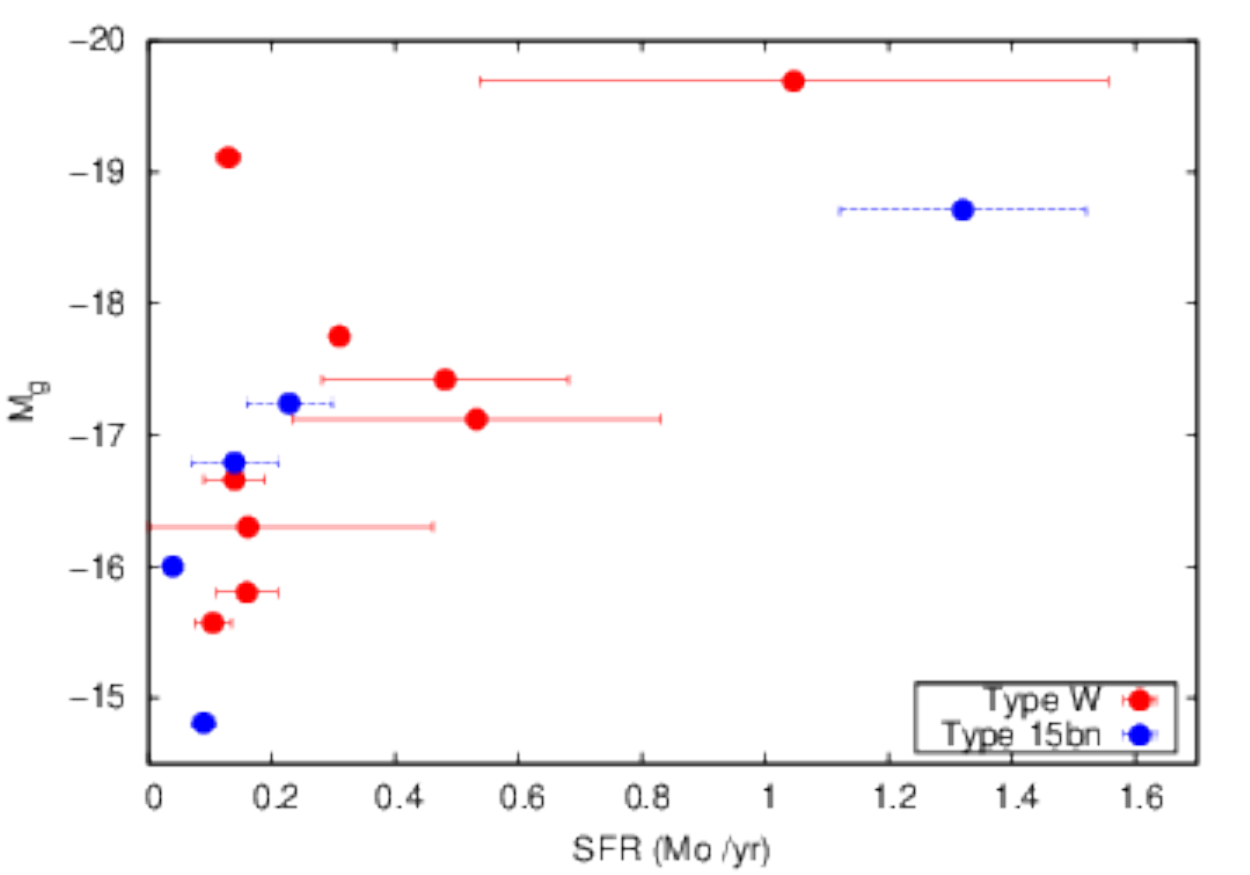}
\includegraphics[width=6cm]{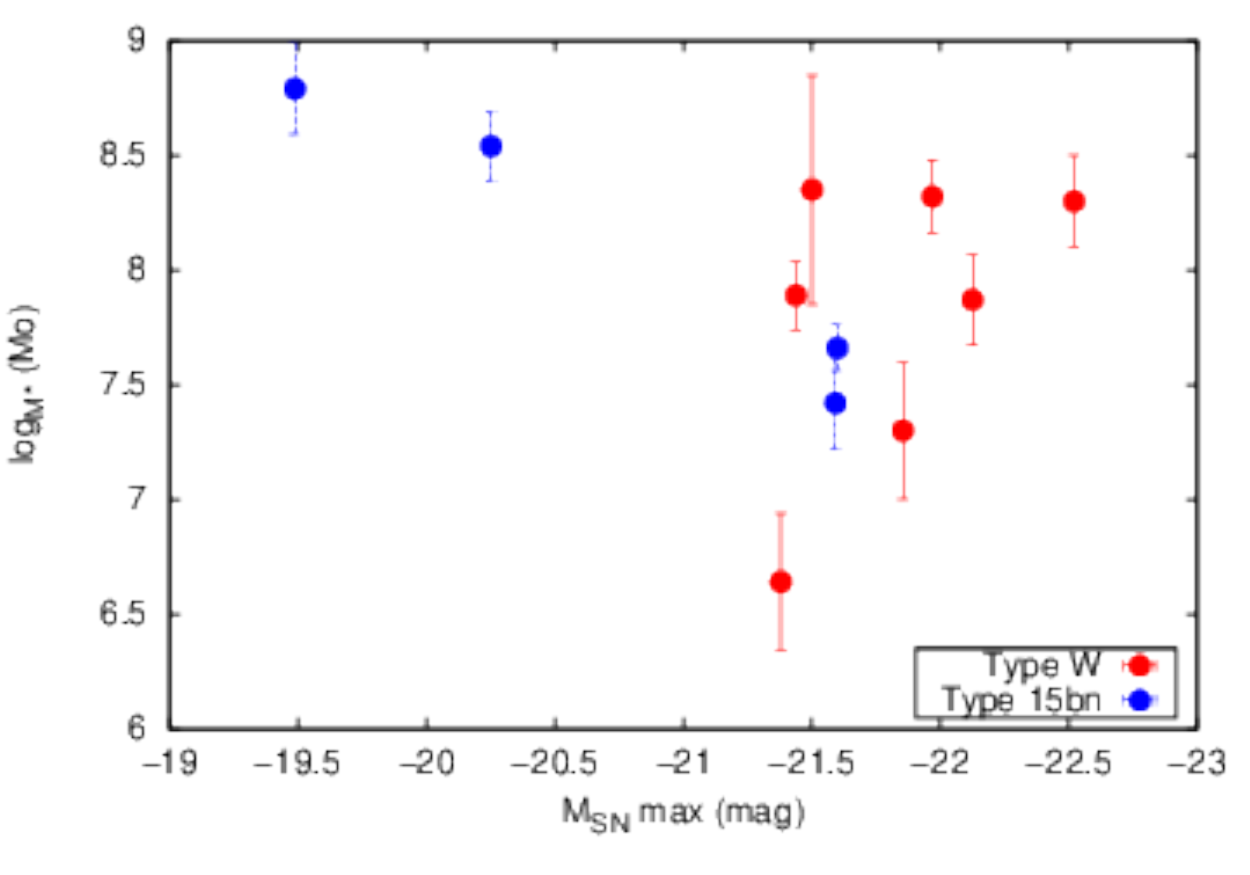}
\caption{Top left: star formation rate vs stellar mass of the host galaxies of the studied SLSNe-I. Top right: star formation rate vs brightness of the galaxy of the SLSNe-I having available data. Bottom: absolute brightness of the supernova in the moment of the maximum ($M_{\rm SN}$ max) vs the stellar mass off the host galaxy. Type W SLSNe-I are plotted with red, while Type 15bn SLSNe-I are shown with blue dots. }
\label{fig:host}
\end{figure*}

\begin{table*}
\caption{Host galaxy properties of the studied Type W and Type 15bn SLSNe-I, where data from the literature were available: their redshift (z), the absolute brightness of the host galaxy in g band (M$_{\rm g}$), the star formation rate (SFR), the logarithm of the stellar mass, the morphology, and the according reference.}
\label{tab:host}
\centering
\begin{tabular}{lcccccc}
\hline
\hline
SN  & z & M$_{\rm g}$ & SFR (err) & log(M*) (err) & Morphology & Reference \\
 & & (mag) & $M_\odot$ yr$^{-1}$  & $M_\odot$  & & \\
\hline
\multicolumn{7}{c}{Type W} \\
\hline
SN~2005ap & 0.283 & -19.11 & 0.13 (0.02) & 7.95 (0.1) & -- &  \citet{schulze18}\\
PTF09atu & 0.5015 & -16.3 & 0.162 (0.3) & 8.35 (0.5) & Faint dwarf & \citet{perley16} \\
PTF09cnd & 0.2584 & -17.42 & 0.481 (0.2) & 8.32 (0.16) & Compact dwarf & \citet{perley16} \\
SN~2010gx & 0.2297 & -17.12 & 0.532 (0.3) & 7.87 (0.2) & Compact dwarf & \citet{perley16} \\
SN~2010kd & 0.101 & -15.57 & 0.105 (0.03) & 7.3 (0.3) & Dwarf &\citet{schulze18} \\
PTF12dam & 0.1073 & -19.69 & 1.046 (0.51) & 8.3 (0.2) & Compact dwarf & \citet{perley16} \\
LSQ14mo & 0.256 & -16.66 & 0.14 (0.05) & 7.89 (0.15) & -- & \citet{schulze18} \\
SN~2018hti & 0.0612 & -17.75 & 0.31 (0.01) & -- (--) & Dwarf & \citet{lin20} \\
\hline
\multicolumn{7}{c}{Type 15bn} \\
\hline
PTF10nmn & 0.1237 & -17.24 & 0.229 (0.07) & 8.79 (0.2) & Irregular & \citet{perley16} \\
LSQ12dlf & 0.255 & -16 & 0.04 (0.001) & 7.66 (0.1) & -- & \citet{schulze18} \\
LSQ14an & 0.163 & -18.71 & 1.32 (0.2) & 8.54 (0.15) & -- & \citet{schulze18} \\
SN~2015bn & 0.110 & -14.81 & 0.09 (0.02) & 7.5 (0.4) & Dwarf & \citet{schulze18} \\
SSS120810 & 0.156 & -16.79 & 0.14 (0.07) & 7.42 (0.2) & -- & \citet{schulze18} \\
\hline
\end{tabular}
\end{table*}

In Figure \ref{fig:host}, the star formation rate vs stellar mass (top left), the star formation rate vs brightness of the galaxy (top right), and the absolute brightness of the supernova in the moment of the maximum ($M_{\rm SN}$ max) vs the stellar mass of the host galaxy (bottom) relations can be seen. Red dots denote to Type W SLSNe-I, while Type 15bn SLSNe-I are marked with blue dots. Figure \ref{fig:host} shows that no separation between Type W and Type 15bn SLSNe-I can be found in the respect of SFR vs log(M*), and neither in case of SFR vs $M_{\rm g}$. Nevertheless, from the bottom panel, it seems that Type 15bn SLSNe-I tend to be fainter in the moment of the maximum compared to Type W objects, and they tend to appear in hosts having larger stellar masses.  However, as the size of the sample is very limited, no firm statements can be concluded from the bottom panel of Figure \ref{fig:host}, since the bimodality cannot undoubtedly be seen between Type W and Type 15bn groups, and this apparent difference may be caused by selection bias. In the future, the examination of a larger set of host galaxy data of SLSNe-I may help in finding differences between the explosion environments of the two sub-types.

\section{Summary}\label{sec:sum}

In this study, we examined the post-maximum spectroscopic properties of a sample of 27 SLSNe-I in order to search for differences between the previously revealed (see \citep{ktr21}) Type W and Type 15bn subgroups. In \citet{ktr22}, we searched for answers behind the pre-maximum bimodality of SLSNe-I by modeling the available pre-maximum spectra of our sample, and found that Type W SLSNe-I tend to show a hotter photosphere compared to Type 15bn SLSNe-I. Furthermore, their early light-curve evolution and geometry may be different as well. Our main goal here was to examine the available post-maximum data of the studied Type W and Type 15bn SLSNe-I for completenessin order to find out if there is a difference between these groups in their later phase evolution or environment as well.

We examined post-maximum data taken at 3 different epochs: an "early" post maximum phase ranging between +0 and +12 days, a "later" post-maximum phase having data in between +10 and +35 days phase, and pseudo-nebular spectra, where the nebular features are clearly visible together with the signs of continuum emission. After modeling the post-maximum the post-maximum spectra of PTF12dam (Type W SLSN-I) and SN~2015bn (Type 15bn SLSN-I), it was found that after the moment of the maximum, the W-shaped absorption feature diminishes from Type W objects, and their ion composition becomes similar to the members of the Type 15bn group. By the pseudo-nebular phase, the two sub-types evolve uniformly, consistently with \citet{nicholl19}.

After calculating the photospheric temperatures and velocities of the objects in the sample, it was revealed that both Type W and Type 15bn SLSNe become cooler than 10000 K, and the photosphere recedes below 10000 km s$^{-1}$ in most cases, thus the two groups become quite similar in this respect as well. One apparent post-maximum difference may be found in the pEW of the place, where the W-shaped feature is/was present in case of Type W SLSNe-I: while the pEW of this feature is the maximal at 13 days in case of PTF12dam (Type W), in case of SN~2015bn, this feature has a maximum strength at +80 days phase. This trend remains true for the other members of the sample, however, in order to strengthen this statement, the more detailed examination of the post-maximum data is needed. The basin-shaped spectral feature between  5290 and 6390  \AA\ shows a strengthening trend in the post-maximum phases in case of both groups, while the pEW of the O I $\lambda$7775 line shows no trend in either. 

Finally, host galaxy properties were examined in case of 14 SLSNe-I in our sample, where the literature had available data, and no significant different was found in the enviromnents of Type W and Type 15bn SLSNe-I. 

In the following enumeration, we summarize the main results regarding the empirical and physical differences and similarities between Type W and Type 15bn SLSNe-I that were examined in \citet{ktr21}, \citet{ktr22}, and the present paper. Table \ref{tab:sum} presents the compact form of our main findings/hypotheses.

\begin{enumerate}
    \item {\bf Empirical properties}: In \citet{ktr21}, while calculating the ejecta masses of a sample containing 28 SLSNe-I, it was revealed that SLSNe-I may be divided into two additional subgroups by the presence/absence of a W-shaped feature, usually modeled as O II. These new subgroups were named Type W and Type 15bn SLSNe-I. Apart from this apparend difference, they were found show bimodality in their spectroscopic evolution: while all Type 15bn SLSNe-I showed spectroscopically slow evolution ($v_{\rm phot} < $  16000 km s$^{-1}$ near maximum), Type W SLSNe-I were present in both in the fast and the slow-evolving spectroscopic sub-types. Regarding photometrical evolution and rise-times, Type W and Type 15bn objects were quite diverse, and showed evolution time-scales from a few 10 days to $\sim$100 days (see Figure 12 in \citet{ktr21}).

    \item {\bf Photospheric temperature and ion composition}: The pre-maximum $T_{\rm phot}$ evolution and chemical composition of our sample was presented in \citet{ktr22}, while the present paper consists the post-maximum examination of these values. It was found that during the pre-maximum phases, Type W SLSNe-I tend to show a hotter photosphere ($T >$ 12000 K) compared to Type 15bn objects (T $<$ 12000 K), and accordingly, their ion composition are different as well. While the pre-maximum spectra of Type W SLSNe-I can be modeled using C II, C III, O I, {\bf O II}, O III, Na I, Si II, Si III, Fe II, Fe III and Co III, Type 15bn objects do not show O II, and can be fitted well using C II, C III, O I, {\bf Mg II}, Si II, {\bf Ca II}, Fe II and Fe III (see Figures 2-5 in \citet{ktr22}). On the contrary, the photospheric temperature and ion composition of the two groups become similar by the post-maximum phases (see e.g. Figure \ref{fig:tph_vs_vph}), as the O II disappears from Type W SLSNe-I with the cooling of the photosphere. 

    \item {\bf Photospheric velocity}: Similarly to the photospheric temperatures, the values of the photospheric velocities tend to be larger in the pre-maximum phase evolution of Type W SLSNe-I (10000 - 23000 km s$^{-1}$), then Type 15bn objects (8000 - 15000 km s$^{1}$; see Figure 7 in \citet{ktr22}), and become similar in the post-maximum phases (see Figure \ref{fig:tph_vs_vph}. 

    \item {\bf Pseudo-equivalent width of several features}: The pEW of 3 regions of the post-maximum spectra were examined for the studied sample. These are at the wavelength of the suspected W-shaped absorption (between 4166 and 5266 \AA), a basin-shaped feature present between  5290 and 6390 \AA, and the region of the O I $\lambda$7775 line (7295 $–$ 7797 \AA).  Differences between Type W and Type 15bn SLSNe-I were only found in the strength of the first mentioned region, while the latter two were similar in case of the two sub-types (see Figures \ref{fig:15bn_ew} and \ref{fig:ew}). 

    \item {\bf Other properties}: Firstly, from the literature, information on the presence/absence of the so-called {\bf early bumps} were searched for in Section 3.2 in \citet{ktr22}, and it was found that 6 objects in the Type W group show light-curve undulations in the earliest detected phases of the evolution, while the light-curve of Type 15bn SLSNe-I showed smooth LC-evolution in the pre-maximum phases, suggesting that these objects may originate from different kind of progenitors, or different environments. Secondly, 5 SLSNe-I in the sample had available {\bf polarimetric data} (see  Section 4.3 \citet{ktr22} for further details). Regarding the available data the following hypothesis was taken: maybe the progenitor of Type W SLSNe-I is spherically symmetric (shows null-polarization), and Type 15bn SLSne-I can be described using a two-layered model consisting of a more symmetric outer layer of C/O, and an asymmetric inner layer built up of heavier elements, implying again that the two sub-types originate from different kind of explosions. However, in the present paper, {\bf properties of the host galaxies} were studied, and it was found that regarding the morphology, the SFR, the stellar mass and the brightness,  the hosts of Type W and Type 15bn SLSNe-I cannot be distinguished from each other. 
 \end{enumerate}

It is concluded that the main difference between Type W and Type 15bn SLSNe-I can be found in their pre-maximum behavior, and by the post-maximum phases, they become nearly homogeneous. In the future, when larger data-sets will be available, it would be interesting to search for further physical differences between Type W and Type 15bn SLSNe-I in order to understand better their nature, explosion mechanism and environment.

\begin{table*}
\caption{Summary of the results obtained in \citet{ktr21} (KT21), \citet{ktr22} (KT22), and the present paper regarding the similarities/differences between Type W and Type 15bn SLSNe-I. }
\label{tab:sum}
\centering
\begin{tabular}{lllll}
\hline
\hline
  & {\bf Type W SLSNe-I }& {\bf Type 15bn SLSNe-I }& {\bf Illustration} & {\bf Different?}\\
\hline 
\multicolumn{5}{l}{{\bf 1. Empirical properties}} \\
\hline
Pre-maximum W-feature & Present & Missing & Figure \ref{fig:tipus} & Yes \\
Spectroscopic evolution & Fast / Slow & Slow & KT21, Figure 12 & Yes \\
Photometric evolution &  Fast / Slow &  Fast / Slow & KT21, Figure & No \\
\hline
\multicolumn{5}{l}{ {\bf 2. Photospheric temperature and ion composition} } \\
\hline
Pre-maximum $T_{\rm phot}$ & 12000 - 20000 K & 8000 - 12000 K & Figure \ref{fig:grad} & Yes \\
Pre-maximum ions & O II present & Mg II, Ca II present & KT22, Figures 2-5 & Yes \\ 
"Later" post-maximum $T_{\rm phot}$ & 6500 - 11000 K & 6000 - 9000 K & Figure \ref{fig:grad}& No \\
"Later" pre-maximum ions & Similar & Similar & Figure \ref{fig:modeling} & No \\
Pseudo-nebular ions & Uniform & Uniform & Figure \ref{fig:nebular} & No \\
\hline
\multicolumn{5}{l}{{\bf 3. Photospheric velocity}} \\
\hline
Pre-maximum $v_{\rm phot}$ & 10000 - 23000 km s$^{-1}$ & 8000 - 15000  km s$^{-1}$ & KT22, Figure 7 & Yes \\
"Later" post-maximum $v_{\rm phot}$ & 6000 - 12000 km s$^{-1}$ & 6000 - 10000 km s$^{-1}$ & Figure \ref{fig:tph_vs_vph} & No \\
\hline
\multicolumn{5}{l}{\bf {4. The pEW of features}} \\
\hline
W-place & Maximum at $\sim$10 days & Maximum at $\sim$80 days phase & Figure \ref{fig:15bn_ew} & Yes \\ 
Basin feature & Increasing pEW & Increasing pEW & Figure \ref{fig:ew}& No \\ 
O I $\lambda$7775 & No trend & No trend & Figure \ref{fig:ew} & No \\
\hline
\multicolumn{5}{l}{ \bf {5. Other properties }} \\
\hline
Pre-peak bump in LC & Present in 6 objects & Present in 0 objects & KT22, Sec. 4.2 & Yes \\ 
Geometry & Null-polarization (?) & Increasing polarization (?) & KT22, Sec 4.3 & Yes? \\
Host galaxy & high SFR, dwarf & high SFR, dwarf & Sec. \ref{sec:host} & No \\ 
\hline
\end{tabular}
\end{table*}

\acknowledgments

This study is supported by the ÚNKP-22-4 New National Excellence Program of the Ministry for Culture and Innovation from the source of the National Research, Development and Innovation Fund. The project is also supported by the NKFIH/OTKA FK-134432 and the NKFIH/OTKA K-142534 grant of the
National Research, Development and Innovation (NRDI) Office of Hungary.

{}

\newpage

\section{Appendix}
\setcounter{table}{0}
\renewcommand{\thetable}{A\arabic{table}}

\setcounter{figure}{0}
\renewcommand{\thefigure}{A\arabic{figure}}

\begin{table*}
\caption{Best-fit local SYN++ parameters of the modeled spectra of PTF12dam and SN~2015bn. The values of $v_{\rm min}$ and $v_{\rm max}$ are given in 1000 km s$^{-1}$, while  $T_{\rm exc}$ is in 1000 K units. }
\label{tab:syn}
\centering
\begin{tabular}{lccccccc}
\hline
Ions & C II &  O I   & Na I & Si II &  Ca II & Fe II &  Fe III  \\
\hline
  \multicolumn{8}{c}{PTF12dam (+7 days phase)} \\
\hline
$\log\tau$ & & -0.4 &&0.0&&-0.3&-0.5 \\
$v_{\rm min}$ && 8.0&&8.0&&8.0&8.0\\
$v_{\rm max}$ && 50.0&&50.0&&50.0&50.0\\
aux & & 2.0&&8.36&&1.0&2.0\\
$T_{\rm exc}$ && 14.0&&14.0&&14.0&8.0 \\
\hline
 \multicolumn{8}{c}{PTF12dam (+14 days phase) } \\
\hline
$\log\tau$ & & -0.4 &&0.0&&-0.3&-0.5 \\
$v_{\rm min}$ && 8.0&&8.0&&8.0&8.0\\
$v_{\rm max}$ && 50.0&&50.0&&50.0&50.0\\
aux & & 2.0&&3.0&&1.0&2.0\\
$T_{\rm exc}$ && 14.0&&14.0&&14.0&8.0 \\
\hline
 \multicolumn{8}{c}{SN~2015bn (+7 days phase) } \\
\hline
$\log\tau$ &-1.5 &0.1&0.6&-0.4&0.0&-0.1&-0.2 \\
$v_{\rm min}$ &10.0&8.0&15.0&8.0&8.0&8.0&8.0 \\
$v_{\rm max}$ &50.0&50.0&50.0&50.0&50.0&50.0&50.0 \\
aux & 3.0&3.0&2.5&6.0&1.0&1.0&1.0 \\
$T_{\rm exc}$& 13.0 &13.0&13.0&13.0&13.0&10.0&10.0 \\
\hline
 \multicolumn{8}{c}{SN~2015bn (+20 days phase) } \\
\hline
$\log\tau$ &-1.5 &0.1&0.6&-0.4&0.0&0.0&-0.2 \\
$v_{\rm min}$ &10.0&8.0&15.0&8.0&8.0&8.0&8.0 \\
$v_{\rm max}$ &50.0&50.0&50.0&50.0&50.0&50.0&50.0 \\
aux & 3.0&3.0&2.5&6.0&1.0&1.0&1.0 \\
$T_{\rm exc}$& 13.0 &13.0&13.0&13.0&13.0&10.0&10.0 \\
\hline
\hline
\end{tabular}
\end{table*}

\end{document}